\documentstyle[12pt]{article}
\begin{document} 
\begin{flushright} {OITS 668}\\
March 1999
\end{flushright}
\vspace*{1cm}

\begin{center}  {\large {\bf CRITICAL BEHAVIOR OF HADRONIC
FLUCTUATIONS AND THE EFFECT OF FINAL-STATE RANDOMIZATION}}
\vskip .75cm
 {\bf  Rudolph C. Hwa\footnote{E-mail
address: hwa@oregon.uoregon.edu} and  Yuanfang Wu\footnote{Permanent
Address:  Institute of Particle Physics, Hua Zhong Normal University,
Wuhan, China}}

 {\bf Institute of Theoretical Science and Department of
Physics\\ University of Oregon, Eugene, OR 97403-5203}
\end{center}

\begin{abstract}
The critical behaviors of quark-hadron phase transition are explored by use of
the Ising model adapted for hadron production.  Various measures involving the
fluctuations of the produced hadrons in bins of various sizes are examined
with the aim of quantifying the clustering properties that are  universal
features of all critical phenomena.  Some of the measures involve wavelet
analysis.  Two of the measures are found to exhibit the canonical power-law
behavior near the critical temperature.  The effect of final-state randomization is
studied by requiring the produced particles to take random walks in the
transverse plane.  It is demonstrated that for the measures considered the
dependence  on the randomization process is weak.  Since temperature is not a
directly measurable variable, the average hadronic density of a portion of each
event is used as the control variable that is measurable.  The event-to-event
fluctuations are taken into account in the study of the dependence of the chosen
measures on that control variable.  Phenomenologically verifiable critical
behaviors are found and are proposed for use as a signature of quark-hadron
phase transition in relativistic heavy-ion collisions.

\end{abstract}

\section{Introduction}
If indeed quark-gluon plasma is to be produced in relativistic heavy-ion
collisions at high energy, such as at BNL-RHIC, there must be a
quark-hadron phase transition (PT) when the plasma becomes cool enough, at
which point hadrons are created and subsequently detected.  The properties of
that PT have not been extensively investigated for a number of reasons. 
Theoretically, the physics is not perturbative, and the fluctuation of hadronic
observables is beyond the scope of feasibility of lattice gauge calculations at
present.  Experimentally, there is the preoccupation with the
conventional main-stream issues, such as the primordial signatures,
hydrodynamical flow, interferometry, etc.   The most serious
obstruction, however, is the common belief, based mostly on an
extrapolation from low-energy nuclear collisions, that a hadron gas
surrounds the plasma and that the final-state interaction in that gas
obliterates all features of the PT, leaving the hadrons with no
significant memory of their birth to be registered at the detector.  One of the
aims of this paper is to investigate the type of hadronic signature that
can survive the final-state interaction and thereby to dispel the myth
that critical behavior of quark-hadron PT is not observable.

If we take statistical physics as a guide, then we should expect the
critical phenomena related to quark-hadron PT to be a vibrant area of
study.  The familiar critical behaviors of conventional systems are, for
example, $C \sim \left|T - T_c \right|^{-\alpha}$ for the heat
capacity of a thermal system, or $\chi \sim \left|T - T_c
\right|^{-\gamma}$ for the susceptibility of a magnetic system, where
$T_c$ is the critical temperature.  Is there any similar behavior for a
quark system in its transition to hadrons?  Can we find an observable
in heavy-ion collisions, call it $\mu$, such that its behavior near the
critical point is
\begin{eqnarray}
\mu \sim \left|T - T_c \right|^{-\kappa} \quad ?
\label{1}
\end{eqnarray}
Will $\mu$ survive the randomization process in the hadron gas, if it
exists?  Since the temperature $T$ is not directly observable in
heavy-ion collisions, is the critical exponent $\kappa$ measurable?  If
not, what is the nature of the critical behavior that is subject to
experimental test?  These are some of the questions that we shall
address and find answers to.

Our investigation cannot be based on nonperturbative QCD, which is
too difficult to implement in treating the problem posed.  In the
absence of a reliable dynamical framework that can substitute for
QCD, we appeal to the lesson already learned in statistical physics in
that the universal properties of the critical phenomena are
independent of the details of the microscopic dynamics.  Only
dimension and symmetry are of most importance.  In the
Ginzburg-Landau theory a universal scaling behavior on multiplicity
fluctuations was derived
\cite{1}, and later verified experimentally in quantum optics at the
threshold of lasing \cite{2}.  That being a mean-field analysis, a more
elaborate investigation that involves the study of spatial fluctuations
near the critical point was subsequently carried out in the framework
of the Ising model \cite{3}.  The result confirms the average behavior
found in \cite{1}, but the work goes beyond the mean-field
approximation by examining the PT process on a 2D lattice in a way
that can best simulate the hadronization process.  In this paper we
continue to adopt the same approach, the only one available short of
QCD, to study the fluctuation of spatial patterns of hadronization
during PT, and to investigate whether the measures chosen can
survive the randomization process of final-state interaction.

Instead of working with factorial moments, which are the measures of
choice in the study of intermittency \cite{4}, we shall consider
elementary wavelet analysis \cite{5} to enhance the effectiveness of
describing clusters of all sizes.  Furthermore, we shall go beyond
intermittency and study   the
 fluctuations of spatial patterns \cite{6,7}.  A
measure that is based on the wavelet coefficients will be
investigated   in
our search for a description of the critical behavior.

Our physical picture of the heavy-ion collision process at high enough
energy and nuclear masses for quark-gluon plasma to be formed is the
conventional one.  After formation, a cylinder of high temperature
plasma that is created undergoes rapid longitudinal expansion and
slower transverse expansion.  The interior is at $T > T_c$ so the
medium stays in the deconfined state during most part of the time
evolution of the system.  We assume that with two flavors the
quark-hadron PT is second order and that it takes place on the surface
of the cylinder where $T$ is lowest, namely $T_c$ or slightly below. 
Thus the geometry of the system undergoing PT is 2-dimensional.  For
a time interval of the order of 1 fm/c, which is roughly the hadronization time for
the formation of a hadron, we map a section of that cylinderical surface to a
square lattice to be studied in the Ising model.  Hadrons are not formed uniformly
on the surface. At different time segments of the evolutionary history the spatial
patterns of hadron formation are different.  Their fluctuations are simulated by
the changing configurations of the Ising lattice.  It is on the basis of this mapping
between the physical surface and the mathematical lattice that we pursue our
extensive investigation of the 2D Ising model in this paper. 

\section{Hadron Density in the Ising Model}

Our first task is to establish a quantitative connection between the
hadrons produced in a heavy-ion collision and the spins on a 2D
lattice in the Ising model, which we use to simulate a second-order
PT.  A number of intermediate steps are needed to make that
connection; they have been discussed in Ref. \cite{3}, which we
summarize here.

The basis of our approach is the Ginzburg-Landau (GL) theory of PT. 
On the one hand, there is the Ising system of spins that can
implement the GL description through a simple dynamical model capable of
generating microscopic fluctuations \cite{8}.  On the other hand,
particle production in the form of photons is well described by the GL
theory, when a laser is operated at the threshold of lasing \cite{9}. 
Our prediction for the scaling behavior associated with hadron
production in the GL formalism \cite{1,10} turns out to be verified
experimentally by photon production in quantum optics \cite{2}.  Thus
we shall assume that the relationship between the GL order
parameter and the hadron density to be the same as that for the
photon density in the laser problem which is rigorously known.

The GL free energy density is 
\begin{eqnarray}
{\cal F}[\phi] = a\left|\phi(z) \right|^2 + b\left|\phi(z) \right|^4 +
c\left|\partial\phi/ \partial z \right|^2 \quad ,
\label{2}
\end{eqnarray}
where $\phi(z)$ is the order parameter for the spatial coordinate at
$z$ in a 2D space.  The parameters $a$, $b$ and $c$ need not concern
us here.  $\phi(z)$ is the link between hadrons and Ising spins.  On
one hand, the hadron density, in analogy to the photon density, is
\begin{eqnarray}
\rho(z) \propto \left|\phi(z) \right|^2 \quad ,
\label{3}
\end{eqnarray}
while on the other hand $\phi(z)$ is the local mean magnetization
field, i.e., 
\begin{eqnarray}
\phi(z) = A^{-1/2}_c \sum_{j\in A_c} s_j \quad ,
\label{4}
\end{eqnarray}
where the sum is over all lattice sites in a cell of area $A_c =
\epsilon^2$.  In the absence of external field the total magnetization
$m_L$ of the whole lattice may be either positive or negative.  We
define $s_j$ to be positive if the spin is aligned along $m_L$.  Thus if
$\sigma_j$ is the lattice spin having values $\pm 1$, so that $m_L =
\sum_{j\in L^2} \sigma_j$, then we define
\begin{eqnarray}
s_j = \mbox{sgn} \left(m_L \right) \sigma_j \quad ,
\label{5}
\end{eqnarray}
where sgn $\left(m_L \right)$ stands for the sign of $m_L$.  This
change of sign when $m_L$ changes sign is irrelevant to the connection
between the GL theory and the Ising model; however, (\ref{5}) is
important in establishing a connection between hadron multiplicity and
Ising spin.  When $T > T_c$, hadronization is inhibited, and $\phi(z)$ is near
zero since the summation of spins in a cell in (\ref{4}) tends to cancel the
site spins in the disordered state.  If it is nonzero due to fluctuations,
hadrons can be formed, but we associate them only with $\phi(z) > 0$,
as we would for the ordered state, $T < T_c$, where hadrons are
formed more copiously.  In order to relate different event of collision with
different configurations of the Ising spins in a uniform way, we must
associate hadron formation with spins aligned with $m_L$, i.e.,
$\sum_j s_j > 0$, not $\sum_j \sigma_j > 0$.  To summarize, the
hadron density in the {\it i}th cell is
\begin{eqnarray} 
\rho_i = \lambda\left|\sum_{j\in A_c(i)} s_j \right|^2
\Theta\left(\sum_j s_j\right)
\quad,
\label{6}
\end{eqnarray} 
where $\Theta$ is the Heavyside function.
The factor $\lambda$ in (\ref{6}) relates the mathematical quantity on the
right side defined on an Ising lattice to the physical quantity of the
number of particles in a cell emitted from the surface of quark-gluon
plasma formed in a heavy-ion collision.  Since $\lambda$ is an
unknown parameter in our attempt to simulate quark-hadron PT by
the Ising model, we shall vary $\lambda$ and regard any results that are
sensitively  dependent on $\lambda$ as unphysical.

In our Monte Carlo simulation of the Ising model we use the Wolff
algorithm \cite{9,11} to calculate the spin configurations for a lattice of
size $L$ = 256, using the Hamiltonian
\begin{eqnarray}
H = - J\sum_{\left<ij \right>} \sigma_i \sigma_j \quad ,
\label{7}
\end{eqnarray}
where the sum is over the nearest neighbors.  We choose the size of a cell to be
$\epsilon = 4$ so the maximum value of $\rho /\lambda$ in the
ordered phase is $\left(4^2\right)^2 = 256$.  Thus $\lambda$ would
have to be quite small to render the multiplicity in a cell to be a
reasonably small number.  Near $T_c$ we expect $\rho$ to vanish in
many cells.  A bin of size $\delta$ consists of $N_c =
\left(\delta/\epsilon\right)^2$ cells, and the whole lattice consists of
$M = \left(L/\delta\right)^2$ bins.  The average density, averaged
over all cells $i$, bins $k$, and configurations $e$, is
\begin{eqnarray}
\left< \rho \right> = {1 \over \cal{N}} \sum^{\cal{N}}_{e = 1}{1 \over
M} \sum^M_{k = 1} {1 \over N_c} \sum^{N_c}_{i = 1} \rho_i \quad ,
\label{8}
\end{eqnarray}
where $\cal{N}$ is the number of simulated configurations.
In the following the results obtained are for ${\cal N} = 10^4$.

The dependence of $\left< \rho \right>$ on $T$ is shown in Fig.\ 1.  It is
evident that a continuous PT occurs at $T_c$ somewhere between 2.0
and 2.5 in units of $J/k_B$.  The precise value of $T_c$ has been
determined in \cite{3} by examining the scaling behavior of the
normalized factorial moments $F_q$.  It is found that $F_q$ behaves
as $M^{\varphi_q}$ only at $T = 2.315$.  Since the critical point is
characterized by the formation of clusters of all sizes in a scale
independent way, we identify the critical temperature at $T_c =
2.315$.  Note that $\left< \rho \right>$ does not vanish for $T > T_c$,
contrary to the behavior of the average magnetization in the limit of
vanishing external field.  The difference is due to our definition of
hadron density $\rho$ in (\ref{6}) as opposed to the average
magnetization being $\left< m_L\right>$.  For heavy-ion collisions it is
reasonable to expect that hadrons can be produced even at $T > T_c$,
though at reduced rate, due to local fluctuations in quark density and
interactions.

For the sake of visualization of the spatial pattern of hadrons formed at  $T_c$,
we show in Fig.\ 2 a particular configuration on the lattice simulated at $T_c$.
Squares of various sizes denote proportionally the cell densities of hadrons. It is
evident that clusters of all sizes are formed. That is the well-known phenomenon
at PT. Here it is the hadron density in a cell that is shown, not just spin up or down
in the Ising problem.

Now, in search for a behavior analogous to (\ref{1}), we focus on the
$T$ dependence of $\left< \rho \right>$ for $T \leq T_c$.  In Fig.\ 3 we
show $\left< \rho \right> - \left< \rho_c \right>$ vs $T_c - T$ in a
log-log plot with $\lambda$ set at 1, $\left< \rho_c \right>$ being
$\left< \rho \right>$ at $T_c$.  For the range of $T$ examined, it is
clear that there exists a power-law behavior
\begin{eqnarray}
\left< \rho \right> - \left< \rho_c \right> \propto \left( T_c -
T\right)^{\eta}
\label{9}
\end{eqnarray}
with
\begin{eqnarray}
\eta = 1.67 \quad .
\label{10}
\end{eqnarray}
For comparison, the susceptibility of a magnetic system has a critical
exponent of $\sim 1.3$.

If the final-state interaction in the hadron gas exists for several
fm/c, long compared to the typical time for resonance decay ($\sim$
1 fm/c), the multiplicity of stable hadrons (in strong interaction)
produced would remain unchanged during that phase, since a
temperature of $\sim$ 140 MeV is not high enough to cause further
production of particles.  Thus the average hadron density is
insensitive to the final-state interaction.  However, (\ref{9}) is not
what is usually regarded as critical behavior, for which the exponent
in a formula such as (\ref{1}) is negative.  Our search for a suitable
measure of critical behavior therefore continues.

\section{Fluctuations of Hadron Density}

Since the fluctuation in particle production is large near the critical
point, as evidenced by Fig.\ 2,  we next consider the fluctuation of $\rho$ from the
average
$\left< \rho \right>$.  Let the average density in bin $k$ of size $\delta$
be
\begin{eqnarray}
\rho_k = \left( {\epsilon  \over  \delta}\right)^2 \sum^{N_c}_{i =
1} \rho_i
\label{11}
\end{eqnarray}
where the sum is over all cells in bin $k$, the total number of which is
$N_c = \left( \delta / \epsilon\right)^2$.  Near $T_c$, $\rho_k$
fluctuates from bin to bin, especially for small $\delta$.    Let us consider the
moments 
\begin{eqnarray}
A_q(M) = \left<\left({\rho_k  \over \left< \rho\right> } \right)^q
\right> = {1 \over {\cal N} M}\sum_{e,k} \left({\rho_k  \over \left<
\rho\right> } \right)^q 
\label{12}
\end{eqnarray}
where $\left< \rho\right>$ is given in (\ref{8}).  By studying the ratio
$\rho_k  / \left<\rho\right> $, we avoid the dependence on the unknown
$\lambda$; moreover, we have $A_q = 1$ at $q = 1$.  A representation of
the properties of $A_q$ is given by
\begin{eqnarray}
J(M) = \left. { d \over dq } A_q\right|_{q = 1} = \left<{\rho_k  \over
\left< \rho\right>}{\rm ln} {\rho_k  \over
\left< \rho\right>}\right> \quad .
\label{13}
\end{eqnarray}
One can also study $q = 2$ and higher powers, but $J(M)$ is sufficient for
our purpose here.

In Fig.\ 4 we show the result of our simulation on $J(M)$.  If $A_q$ had a
power-law dependence on $M = \left(L/\delta \right)^2$, then $J(M)$
would depend linearly on ln $M$.  Evidently, Fig.\ 4 does not exhibit
any linear dependence at any $T$.  However, the behaviors at various
$T \leq T_c$ are similar.  The situation is reminiscent of the behavior of
$F_q$ in the GL theory \cite{1}, where strict scaling $F_q \propto
M^{\varphi _q}$ is not valid, but  $F_q \propto F_2^{\beta _q}$ is valid. 
Similarly, we can consider here the dependence of $J(M)$ on $J_c(M)$,
where $J_c(M) = J(M, T = T_c)$.  As can be seen from Fig.\ 5, the points for
$T < T_c$ can be well fitted by straight lines.  Let us therefore write
\begin{eqnarray}
J(M, T) = \alpha (T)J_c(M) \quad . 
\label{14}
\end{eqnarray}
This is the property of a generalized form of scaling \cite{12,13}
\begin{eqnarray}
A_q(M, T) \propto g(M)^{\tau_q(T)} \quad ,
\label{15}
\end{eqnarray}
where $g(M)$ is some function of $M$ independent of $T$.  It follows
directly from (\ref{13}) that
\begin{eqnarray}
\alpha (T) = \tau^{\prime}(T)/\tau^{\prime}\left(T_c\right) \quad ,
\label{16}
\end{eqnarray}
where $\left.\tau^{\prime}(T) \equiv d \tau_q(T)/dq \right|_{q = 1} $.  We
do not pursue the study of the generalized scaling (\ref{15}) here.

From the slopes of the straightline fits in Fig.\ 5, we can exhibit the $T$
dependence of $\alpha (T)$, shown in Fig.\ 6. We display $\alpha (T)$ in log-log
plot so as to explore the possibility of the critical behavior
\begin{eqnarray}
\alpha (T) \propto \left( T_c - T\right)^{-\zeta} \quad .
\label{17}
\end{eqnarray}
We have no illusion that (\ref{17}) could be valid in the limit $T
\rightarrow T_c$, since $\alpha (T_c) = 1$, by definition.  However, for a
range of $T$ below $T_c$, Fig.\ 6 does show an approximate linearity.  The
corresponding value of the critical exponent $\eta$ is
\begin{eqnarray}
\zeta = 1.88 \quad .
\label{18}
\end{eqnarray}
Since (\ref{17}) is invalid in the immediate neighborhood of $T_c$, it
cannot be taken seriously as the defining characteristics of quark-hadron
PT.  Instead of (\ref{17}), we regard Fig.\ 6 as containing more information
about the fluctuations of the hadron density near the critical point.  How
such a behavior can be checked experimentally will be discussed in a later
section below.

\section{Fluctuations in Clustering}

In the previous section we have studied the fluctuation properties of
hadron density from bin to bin for all events.  Since the bin density
$\rho_b$ is an average of the cell densities in a bin, it is not sensitive to
the fluctuations of the cell densities within a bin.  Clustering being such a
characteristic feature of critical behavior where clusters of all sizes can
occur, we need to go beyond $\rho_b$ to capture the fluctuations of
multiplicities among neighboring cells within the same bin.  To that end
we shall work with multiplicities instead of densities of hadrons, not only
for calculating spatial fluctuations, but also for considering temporal
fluctuations in a later section when randomization due to final-state
interaction is taken up.

In terms of the cell density $\rho_i$ defined in (\ref{6}) the cell
multiplicity $n_i$ is simply $\epsilon^2 \rho_i$.  Since $\lambda$ in
(\ref{6}) is unspecified, let us absorb the factor $\epsilon^2$ into
$\lambda$, i.e. redefining $\epsilon^2\lambda \rightarrow \lambda$, and
write 
\begin{eqnarray}
n_i = I \left[\lambda \left| \sum_{j\in A_i}s_j\right|^2 \Theta
\left( \sum_{j}s_j\right)\right]\quad ,
\label{19}
\end{eqnarray} 
where $I[x]$ signifies the largest integer $\leq x$.  Thus $n_i$ is discrete,
although $\lambda$ is treated as a continuous parameter $\leq 1$.  For
every configuration simulated on the lattice we have a matrix $C\left(c_1,
c_2 \right)$ that gives the value of the cell multiplicity $n_i$ at the cell
position $\left( c_1, c_2\right)$ in the 2D space, where $c_i$ ranges from 1
to $\ell = L/\epsilon$; i.e., $C\left(c_1, c_2 \right) = n_i \left[i = \left(c_1,
c_2 \right) \right]$.  Figure 2 is a pictorial representation of a possible
configuration whose numerical description would be a particular matrix $C\left(
c_1, c_2\right)$.

Now, for every bin in that $\ell \times \ell$ space we consider a
transform through a function $\psi^H$  that has the following property. 
Divide a square bin into four quadrants:  in the two diagonal quadrants
$\psi^H = + 1$, in the two off-diagonal quadrants $\psi^H = - 1$.  If we use
$x$ and $y$ to denote the two orthogonal coordinates of the square bin,
normalized to unit length on both sides, then the precise definition of
$\psi^H (x,y)$ is
\begin{eqnarray}
\psi^H (x,y) = 1,\quad &0 \leq x  < 1/2, &0 \leq y < 1/2,\nonumber\\ 
= -1,&1/2\leq x  < 1,&0 \leq y  < 1/2,\nonumber\\ 
= -1,&0 \leq x  < 1/2,&1/2 \leq y  < 1,\\ 
= 1,&1/2 \leq x  < 1,&1/2 \leq y  < 1,\nonumber\\ 
= 0,&{\rm elsewhere} \quad
.\nonumber
\label{20}
\end{eqnarray}
This is essentially a 2D Haar wavelet \cite{5}, but any prior knowledge
about wavelet analysis is not needed for our discussion here.

We label a specific bin in the $\ell \times \ell$ space of cells by a triplet
$\left(j, k_1, k_2\right)$, where the bin size $\delta_j$ is 
\begin{eqnarray}
\delta_j = \ell\, 2^{-j}
\label{21}
\end{eqnarray}
and the location of the bin is specified by $\left(k_1, k_2\right)$, with
$k_i$ ranging from 0 to $2^j - 1$.  Clearly, the total number of bins is
$2^{2j}$.  Thus $j$ gives the level of zooming, and $\left(k_1, k_2\right)$
describes the shift at that level.  The label $j$ used here and in the following
should not be confused with the symbol $j$ for site position on the lattice in the
equations from (4) to (7).  We can now define our basic wavelet function in the
$\ell^2$ space
\begin{eqnarray}
\psi_{j, k_1, k_2} \left(c_1, c_2 \right)  = \psi^H \left({c_1  \over 
\delta_j} - k_1 \, , {c_2  \over 
\delta_j} - k_2\right) \quad .
\label{22}
\end{eqnarray}
It vanishes everywhere except in the $j$th level bin located at $\left(k_1,
k_2\right)$, and in that bin it can only be $\pm 1$ depending on the
quadrant that $c_1$ and $c_2$ fall into.

Using $\psi_{j, k_1, k_2} \left(c_1, c_2 \right)$ we define the wavelet
transform of a configuration of cell multiplicities
\begin{eqnarray}
w_{j, k_1, k_2} = \sum_{c_1, c_2} \psi_{j, k_1, k_2} \left(c_1, c_2
\right) C \left(c_1, c_2 \right) \quad ,
\label{23}
\end{eqnarray}
where the sum is over all cells.  It is clear that if $C \left(c_1, c_2 \right)$
is uniform in a bin $\left(j, k_1, k_2\right)$, then the wavelet coefficient
for that bin, $w_{j, k_1, k_2}$, is zero, although the average density
$\rho_b$ for that bin is not.  Thus $w_{j, k_1, k_2}$ measures the
fluctuation from the average in a specific way.  By varying $j$ we can
learn a great deal about the landscape of $C \left(c_1, c_2 \right)$.

To study the fluctuations of the clustering within each configuration, but averaging
over all configurations, we investigate the scaling properties of the moments of the
wavelet coefficients, $w_{j, k_1, k_2}$.
The wavelet coefficients, being more sensitive
to fluctuations than $F_q$, have a better chance of generating a
discernible signature of clustering patterns formed near the critical point.
  To that end, we define for a fixed $j$, for
which $M = 2^{2j}$,
\begin{eqnarray}
B_q(M) = \left<\left({w_{j, k_1, k_2}  \over  \left< w\right>}
\right)^q\right>,
\label{24}
\end{eqnarray}
where
\begin{eqnarray}
\left<\cdots \right> = {1 \over \cal{N}} \sum^{\cal{N}}_{e=1} {1 \over
M}\sum^{2^j - 1}_{k_{1,2}=0}\cdots.
\label{25}
\end{eqnarray}
Furthermore, we have
\begin{eqnarray}
K(M) =\left.  {d  \over dq } B_q\, \right | _{q = 1} = \left<{w_{j, k_1, k_2} 
\over  \left< w\right>}\, {\rm ln}\, {w_{j, k_1, k_2}  \over  \left< w\right>}
\right>,
\label{26}
\end{eqnarray}
which is the quantity that we shall investigate.  In studying the ratio
$w_{j, k_1, k_2}/\left< w\right>$, the effects of $\lambda$ in (\ref{19}) 
tend to cancel, but not completely because $n_i$  is a discretized version of
a noninteger quantity at each cell.  Thus some dependence of $K(M)$  on
$\lambda$ is expected, a property that will be examined.

The erraticity analysis \cite{6,7} originally designed to describe the
event-to-event fluctuations was done using factorial moments $F_q$. 
For the problem of PT  it was found that the erraticity indices turn out to
be very small (of order $10^{-3}$) \cite{12}, smaller than what experimental
data can probably resolve.  What is proposed in (\ref{24})-(\ref{26}) is not
erraticity analysis, and is not aimed at studying the fluctuations from
configuration-to-configuration. Instead, $B_q(M)$ and $K(M)$ are measures of the
fluctuations from bin to bin that are averaged over all configurations. 

In Fig.\ 7 we show $K(M)$ vs
ln $M$ for a number of values of $T \leq T_c$.  The dependence on $M$ is
evidently far more severe than that of $J(M)$ at large $M$.  Due to the fact that
$j$ must be an integer, we cannot study at smaller steps in $M$.  The fluctuations
in small bins are clearly very strong, and they do not vanish in large bins,
lest $K(M)$ would approach zero, as $M \rightarrow 1$.  It means that
hadrons are formed in clusters of all sizes, roughly with as much empty
space on the lattice (unhadronized quarks in real space) between the
clusters as there are hadronized cells.

To facilitate a description of the $T$ dependence of $K(M,T)$, we plot
$K(M,T)$ vs $K_c(M) = K(M,T_c)$, varying $M$ parametrically, as shown in
Fig.\ 8.  Despite our inability to add points corresponding to intermediate
bin sizes, the linearity of the behavior is self-evident.  The points can be
well fitted by straight lines, yielding the form
\begin{eqnarray}
K(M,T) = \beta_0(T) + \beta(T)K_c(M) \quad.
\label{27}
\end{eqnarray}
Thus we can extract the slope $\beta(T)$ that depends only on $T$, and
not on the bin sizes.  It therefore characterizes the intrinsic properties of
the system under PT, and may be more general than the Ising model
used to derive it.

Figures 7 and 8 show the results of the calculation where $\lambda$ has
been set to 1 in (\ref{19}).  We now decrease $\lambda$ to see how
$\beta(T)$ depends on $\lambda$.  That is shown in Fig.\ 9.  Considering
the fact that $\lambda$ has been varied over two orders of magnitude,
the form of $\beta(T)$ is basically stable.  If we fit the last three points
by straight lines in accordance to the formula
\begin{eqnarray}
\beta(T) \propto \left(T_c - T\right)^{-\kappa} \quad ,
\label{28}
\end{eqnarray}
the value of $\kappa$ varies weakly with $\lambda$, as shown in Fig.\ 10,
except at $\lambda = 0.02$ and $0.03$ for unimportant reasons having to do
with the discretization of $n_i$.  Hereafter we shall adopt $\lambda =
0.05$ as the representative value and set
\begin{eqnarray}
\kappa = 2.2 \quad .
\label{29}
\end{eqnarray}
Although (\ref{28}) is not valid as $T \rightarrow T_c$, since by definition
$\beta \left(T_c \right) = 1$, its approximate validity in a narrow range of
$T$ nevertheless captures a behavior remarkably similar to the
conventional critical behaviors of familiar thermal and magnetic
observables.

Since $\alpha(T)$ and $\beta(T)$ look very similar in Figs.\ 6 and 9, it is sensible
to relate them directly, using $T$ only as an implicit  variable. That is, we plot
$\beta$ against $\alpha$, as in Fig.\ 11, where all of the five temperature
points  are shown.  They correspond to   the $\lambda=0.05$ points of
$\alpha(T)$ and $\beta(T)$ in Figs.\ 6 and 9.
 We see that there
exists a  linear relationship, as exhibited by the straightline fit
\begin{eqnarray}
\beta = a+b\alpha    \quad,
\label{30}
\end{eqnarray} 
where $a= -0.0623$ and $ b= 1.11$.  It is particularly remarkable in view of the
nonlinear behavior in Figs.\ 6 and 9.  There has been no prior suspicion that such a
relationship should exist, so the experimental verification of (\ref{30}) would be
of great interest.

	One may wonder how (\ref{30}) can be consistent with (\ref{17}) and (\ref{28})
that have different exponents. Indeed, if one examines the linear fits in the log-log
plots in Figs.\ 6 and 9, the deviations from the straightlines are larger than the
deviations of the points in Fig.\ 11 from the solid line in that figure.  In other
words, when discrepencies from best fits are taken into account, the linear
behavior exhibited in Fig.\ 11 is a better fit of the dependence of $\beta$ on
$\alpha$, which is especially significant in view of the full range of all five points
being related.

In the  discussion in connection with (\ref{17}) and (\ref{28}) we have
emphasized the behaviors of $\alpha$ and
$\beta$ in the region where $T$ is not in the immediate neighborhood of $T_c$. 
We now examine further the region where $T$ is  closer to $T_c$.  
In Fig.\ 12 we have plotted the curves for $K(M,T)$ vs $J(M,T)$ for $T \leq T_c$.
The solid curve for $T=T_c$ is of particular significance if the PT occurs only at
$T_c$.  It is evident that the curves for $T$  close to $T_c$ have very similar
shapes. Thus it appears that by shifting those curves they can all be overlaid on a
universal curve that coincides with the curve at $T_c$. Since shifting horizontally
and vertically correspond to rescaling the horizontal and vertical variables, we
have replotted the curves in a set of new variables that involve changes in
scales.  The result is shown in Fig.\ 13, where $K(T)(T/T_c)^2$ is plotted against
$J(T)(T/T_c)^{-10}$, with $M$ being varied implicitly.  All the curves for
$2.28\leq T\leq T_c=2.315$ overlap and form a ``universal" curve.  Such a scaling
behavior involving a range of $T$  may be
of theoretical interest only at this point, since $T$ is not measurable. 
Nevertheless, the existence of such a universal curve is suggestive of some
property that is like the Kadanoff scaling, where a change in the size of spin
blocks can lead to a shift of the effective temperature \cite{15}.  Of course, the
quantities involved here are quite different, and the reason for the universality
found here deserves further investigation.

\section{Observable Consequences}

The results that we have obtained so far from analyzing the data
generated by the Ising model are the $T$ dependences of $\left<\rho
\right>$, $\alpha(T)$ and $\beta(T)$.  Such results have no direct
phenomenological implications, since the temperature $T$ is not a directly
measurable quantity in heavy-ion collisions, let alone the precise value
of $T_c$.  Thus equations such as (\ref{9}), (\ref{17}) and (\ref{28}) are
primarily of theoretical interest.  They are the results of a search for
behaviors that resembles the conventional critical behaviors.  Now we ask,
given these theoretical hints, what observable consequences of PT can be
suggested that are experimentally testable.

To proceed, it is necessary to recognize first the important aspects in
which the realistic situation of quark-hadron PT in heavy-ion collisions is
more complicated than what has been simulated on the Ising lattice. A
lattice configuration generated by the Ising model can represent  the
hadronization pattern on the surface of the plasma cylinder for only the
time duration of the hadronization time $t_h$, roughly 1 fm/c.  As time
proceeds, the pattern changes, just as the lattice configuration changes
upon repeated simulation.  Under the assumption that the plasma interior
is hot, with $T > T_c$, phase transition occurs on the surface at $T \leq
T_c$ over a period of time long compared to $t_h$.  What needs to be
analyzed in the heavy-ion experiment is the hadronization patterns
separated from one another in time segments of about $t_h$ long.  While
that may be hard to achieve in present experiments, one can approximate
that by making $p_T$ cuts in small $\Delta p_T$ intervals, since one
expects from hydrodynamical considerations that a correspondence
between $p_T$ and evolution time $\tau$ exists such that hadrons
emitted early have higher $p_T$ than those emitted later \cite{14}.  We
proceed under the supposition that it is possible to obtain from the
heavy-ion data collected at a fixed transverse energy $E_T$ many
samples consisting of portions of the events that correspond to the hadronic
patterns in the $\eta$-$\phi$ space and are the realistic counterparts of the
simulated hadronic configurations
$C\left(c_1, c_2\right)$ on the Ising lattice.

In the Ising model we can adjust the temperature $T$ and study the $T$
dependences of various measures.  In heavy-ion collisions that cannot be
done.  We then consider two possible scenarios.  First, we assume that the
quark-hadron $PT$ occurs at $T_c$ only, and no other $T$.  Second, we
consider the possibility that various conditions of hydrodynamical flow
may lead to hadronization at various $T$ below $T_c$ at different points
of time in the evolutionary history of the plasma.  Let us refer to these
two scenarios as A and B, respectively.

In scenario A we give up the possibility of studying the $T$  dependence
of PT.  Our analysis of the data generated on the lattice nevertheless can
serve as a guide to the analysis that can be done on the heavy-ion data. 
Since $J_c(M)$ and $K_c(M)$ are both observable, their dependences on
$M$ in Figs.\ 4 and 7 can both be checked.  Due to the absence of an
absolute relationship between bin sizes on the lattice and those in the
experiments, it is best to examine the dependence of $K_c$ on $J_c$ with $M$
being parametrically varied.  Such a dependence  is already shown
by the solid line in Fig.\ 12  and by Fig.\ 13.
Clearly, such a behavior is independent of the definitions of bin sizes on the lattice
or in the experiment, and is a definitive characterization of the fluctuations of
hadronic clusters at PT. The verification of the relationship bewteen $K_c$ and
$J_c$ would imply that the hadrons detected are created by a quark-hadron PT at
$T_c$. 

 In scenario B where quark-hadron PT can occur at a range of
$T$, the problem is significantly more complicated and will be the subject of
our discussion for the remainder of this section.  Let us first define a sample to be
a portion of an event obtained by certain cuts (such as by a narrow $\Delta p_T$
cut discussed in the beginning of this section) so that a sample corresponds to a
configuration simulated on the Ising lattice.  We assume that a sample is the
result of hadronization at a common $T$, which may differ from $T_c$.
 That $T$ may
vary from sample to sample.   For every sample the average density $\bar\rho$
of the hadronic configuration  can be determined experimentally and should
correspond to the average density   calculated on the lattice according to
\begin{eqnarray}
\bar\rho = {1 \over
M} \sum^M_{k = 1} {1 \over N_c} \sum^{N_c}_{i = 1} \rho_i \quad ,
\label{31}
\end{eqnarray}
 This
$\bar\rho$ is not the same as the theoretical overall average density
$\left<\rho\right>$ defined in (\ref{8}), since the latter is averaged over all
configurations at the same $T$, 
whereas $\bar\rho$ refers to a specific configuration on the lattice or a specific
sample in the experiment.
  Since $\bar\rho$ can vary from sample to sample
(as they do from configuration to configuration on the lattice) even when all
samples are at the same temperature, let us first examine the distribution  
$P(\bar\rho,T)$ of
$\bar\rho$ at fixed $T$.  We have calculated that distribution at three
representive values of $T$ with the result shown in Fig.\ 14.   As $T$ is
decreased from $T_c$, the average density increases, as more hadrons are
produced in a phase transition.  The shapes of the distributions vary quite
significantly as functions of $T$.  On the lattice the cell
densities are bounded by  a maximum, according to (\ref{6}), of $\rho_{i,{\rm
max}}=256\lambda=12.8$ (for $\lambda=0.05$).   The normalization of $\bar\rho$
should not be taken seriously in phenomenology, since the density on a lattice
cannot be rigorously related to the hadron density in an experiment.  We should
therefore regard $\bar\rho$ as having an arbitrary scale.  Nonetheless, these
distributions at fixed
$T$ are important evidences that the event-to-event (and therefore
configuration-to-configuration) fluctuations are large even if all kinematical
variables of the collision processes are fixed, and if the uncontrollable
dynamical variable $T$ at hadronization is held constant.

In view of the rather wide distributions in
Fig.\ 14 (which are all normalized to 1), it is clear that any experimental measure
categorized according to the observed
$\bar\rho$ must involve a convolution over the unobserved $T$. Thus, for
example, in studying $K(M)$ we should convert its $T$ dependence to $\bar\rho$
dependence by use of the formula
\begin{eqnarray}
K(M,\bar\rho) = \int dT\ P(\bar\rho, T)\ Q(T)\ K(M,T)\quad,
\label{32}
\end{eqnarray}
where $Q(T)$ is the probability that the quark-hadron PT will take place at
$T\leq T_c$.

We have no deep  insight on what $Q(T)$ should be. It can depend on the
thermodynamics and hydrodynamics of the heavy-ion collision problem.  One
would expect that PT should take place in the immediate neighborhood of $T_c$. 
We shall take an exponential form for $Q(T)$ as a working hypothesis. Since the
calculations in Secs.\ 3 and 4 are done for the range of $T$ between $T_c=2.315$
and $T=2.25$ and interesting critical behaviors have been found theoretically in
that range, we shall proceed on the assumption that $Q(T)$ is important only in
that range.  Accordingly, we adopt the following form
\begin{eqnarray}
Q(T) = q_0\ {\rm exp}\, [-(T_c-T)/0.08], \qquad 2.24\leq T \leq 2.32,
\label{33}
\end{eqnarray}
and require that it is zero elsewhere.  The normalization factor is $q_0=[0.08
(1-e^{-1})]^{-1}$.

We now can redo the calculations on $J(M)$ and $K(M)$ as in Secs.\ 3 and 4
 by use of the convolution as in (\ref{32}) and present the results in terms of
$\bar\rho$. We shall set $\lambda=0.05$.  In Fig.\ 15 we show $J(M,\bar\rho)$
for various bins of values of $\bar\rho$.  Note that the highest curve (in solid
line) is for
$3.75<\bar\rho<4.0$, for which we shall use $\bar\rho_0=3.88$ to denote that
bin, a value that shall play the role of $T_c$ in Fig.\ 3.  At lower $\bar\rho$ the
curve for $J(M)$ is lower; this fringe effect shall be ignored in the scaling analysis
 below. That is, we shall consider only $\bar\rho>\bar\rho_0$ in the following.
The similarity of the curves suggest following  the same procedure as in Sec.\ 3.
Thus we define $J_0(M)=J(M,\bar\rho=\bar\rho_0)$, and examine
$J(M,\bar\rho)$ vs $J_0(M)$.  The result is shown in Fig.\ 16.  The linear
behavior is remarkable. Defining the slope to be $\alpha(\bar\rho)$, we show its
$\bar\rho$ dependence in Fig.\ 17. A power-law behavior can be identified, as
indicated by the straight line, yielding
\begin{eqnarray}
\alpha(\bar\rho) \propto (\bar\rho-\bar\rho_0)^{-\bar\zeta}, 
\qquad \bar\zeta= 0.97.   \label{34}
\end{eqnarray}
 
The same procedure can be applied to $K(M,\bar\rho)$, as done in Figs.\ 18-20.
With $K_0(M)=K(M,\bar\rho_0)$ the linearity of the lines in Fig.\ 19 is even better
than that in Fig.\ 7. Again, defining the slope to be $\beta(\bar\rho)$, we obtain
from Fig.\ 20 the behavior
\begin{eqnarray}
\beta(\bar\rho) \propto (\bar\rho-\bar\rho_0)^{-\bar\kappa}, 
\qquad \bar\kappa= 0.97.   \label{35}
\end{eqnarray}
Evidently, the behavior of $\beta(\bar\rho)$ is essentially identical to that of
$\alpha(\bar\rho)$.  Equations (\ref{34}) and (\ref{35}) represent the closest that
we can identify as the critical behavior of quark-hadron PT. This time all the
quantities involved are experimentally measurable.

Finally, we can plot $\beta(\bar\rho)$ vs $\alpha(\bar\rho)$ with $\bar\rho$
being the parametric variable. All five points fall on a straight line in Fig.\ 21. The
linear behavior can be well expressed by the amazing formula
\begin{eqnarray}
 \alpha(\bar\rho) = \beta(\bar\rho)
\label{35.1}
\end{eqnarray}
with unit slope.   We have no explanation for the simplicity of (\ref{35.1}).  It
would be highly significant if this behavior can be verified by experiment.

\section{Effects of Randomization}

What we have studied so far are the properties of the hadrons formed on
the cylindrical surface, which is mapped onto the Ising lattice.  The
observed hadrons that reach the detector must traverse the space
between the cylindrical surface and the detector.  If a hadron gas
surrounds the plasma cylinder, then a hadron formed on the surface must
undergo final-state interactions with other hadrons before it can move as a
free particle to the detector.  Assuming no further particle production in
this last phase, the hadrons can only shift their positions in the $\eta$-$\phi$
plane in a random way, since the hadron gas plays the role of random scatterers
and can offer no organized forces on a traversing hadron.

We shall take into account this randomization process by requiring each
hadron to take random walks on the lattice.  The step size is the distance
between neighboring cells, i.e., $\epsilon$.  If a cell has $n_i >1$, then each
particle in the cell takes steps that are independent of what other
particles in the cell do.  A particle has five possible positions in each random
step, four neighboring cells plus the original cell for no transverse movement. 
Denoting the number of steps by $\nu$, we consider $\nu$ ranging from 1 to 6,
although it is generally believed that 3-4 steps are adequate to represent
the effect of randomization by the interactions in a hadron gas.

In considering random walk we must focus on the whereabout of the
particles, so hadron density must be converted to particle number by a
specific choice of $\lambda$.  Although we have avoided that choice in Sec.
3, we have found in Sec. 4 that $\lambda = 0.05$ is an appropriate choice. 
Thus we use the same $\lambda$ now in calculating the effect of random
walk on $J(M, T)$ and then on $\alpha (T)$.  Fig.\ 22(a) summarizes the
effect for $\nu$ up to 4, where we stop to avoid overlap of the points on the
figure at higher
$\nu$.  The general shape of $\alpha (T)$ is unchanged.  Quantitative dependence
of the index $\zeta$, defined in (\ref{17}), on $\nu$ is shown in Fig.\ 22(b),
exhibiting insensitivity to randomization.

The same analysis has been done for $K(M,T)$ and then $\beta(T)$ with
the result shown in Fig.\ 23.  In this case the effect of random walk is more
pronounced, especially at low $\nu$, but for $4 \leq \nu \leq 6$ the effect
stabilizes.  Although $\beta(T)$ suffers more change than $\alpha (T)$, the
overall effect does not invalidate the usefulness of the measures that we
have considered.  We show in Fig.\ 24 the effect of $\nu$ steps of
randomization on the dependence of $\beta$ on $\alpha$.  Evidently, the
general linear dependence is unaltered by the randomization process.

We shall not redo the above calculation for  $\alpha$ and $\beta$ as functions of
$\bar\rho$, since the effect of randomization should be similar  on
the experimental $\alpha(\bar\rho)$ and
$\beta(\bar\rho)$.  If the experimental data can produce  a figure such as Fig.\
21, showing that  $\beta(\bar\rho)$ is essentially the same as
$\alpha(\bar\rho)$, we can conclude on the basis of the study considered here
that the effect of randomization is not so severe as to render the whole approach
ineffective.  We infer that the measures
$J(M)$ and
$K(M)$ must not be very sensitive to the redistribution of particles in the
transverse plane, since it is not the individual local property of a hadron but the
$\it {scaling}$ properties of clusters of hadrons over a wide range of bin sizes that
the fluctuation measures have extracted.

\section{Summary} 
Using the Ising model and defining hadrons on the lattice in accordance to the
Ginsburg-Landau description of quark-hadron PT, we have found a number of
interesting features about the critical behavior.  First of all, the average hadron
density on the surface of the plasma cylinder for a time interval around the
hadronization time of a hadron depends on the surface temperature according as 
\begin{eqnarray}
\left<\rho\right> - \left<\rho_c\right> \propto (T_c-T)^{\eta}\quad, \qquad
\eta=1.67\quad. \label{36}
\end{eqnarray}
Then there are two quantities $\alpha(T)$ and $\beta(T)$, defined in (\ref{14})
and (\ref{27}), that behave as
\begin{eqnarray}
\alpha(T) \propto (T_c-T)^{-\zeta}\quad, \qquad
\zeta=1.88\quad. \label{37}\\
\beta(T) \propto (T_c-T)^{-\kappa}\quad, \qquad
\kappa=2.2\quad. \label{38}
\end{eqnarray}
These are behaviors of theoretical interest, since $T$ is not directly measurable. 
Nonetheless, (\ref{37}) and (\ref{38}) confirm that there exist measures in the
hadronic variables that exhibit the canonical form of critical behaviors.

Since the average density $\bar\rho$ in a sample is measurable, 
we have shown how the critical behaviors can be reexpressed in terms of
$\bar\rho$ as
\begin{eqnarray}
\alpha(\bar\rho) \propto (\bar\rho-\bar\rho_0)^{-\bar\zeta}, \qquad\quad
\beta(\bar\rho) \propto (\bar\rho-\bar\rho_0)^{-\bar\kappa}, \label{39}
\end{eqnarray}
\begin{eqnarray}
\bar\zeta=\bar\kappa=0.97\quad. \label{40}
\end{eqnarray}
The equations in (\ref{39}) are expressed in terms of observables
and should  be checked by experiments.
Furthermore, the direct dependence of $\beta(\bar\rho)$ on
$\alpha(\bar\rho)$ over a wider range of $\bar\rho$ can also be examined.  If it
turns out   that  they are related approximately by $\alpha(\bar\rho) \approx
\beta(\bar\rho)$, then that would be a less restrictive, yet nevertheless valid,
experimental signature of the critical behavior of a quark-hadron PT.

 Experimental study of $J(M)$ and $K(M)$, defined in (\ref{13}) and (\ref{26}), 
involves less analysis and thus are more reliable phenomenologically. If PT occurs
at
$T=T_c$ only, then $J_c(M)$ and $K_c(M)$ are directly measurable. The
verification of $K_c$ vs $J_c$, as shown by the solid line in Fig.\ 12, would
strongly indicate the occurrence of a PT.

None of the phenomenology suggested above would mean anything realistic if the
hadron gas in the final state erases all memory of the properties of phase
transition before the particles reach the detector.  We have studied the effect of
the randomization due to hadron gas on our measures, and found that within
the framework of our modeling none of them are seriously influenced by the
randomization procedure so long as the number of random steps in the
$\eta$-$\phi$ plane is not too large.  Those measures involve the fluctuations
from bin to bin of either the average density of hadrons or the wavelet coefficients
of the spatial patterns, both of which are evidently insensitive to the
randomization.  

An important assumption is made in this work, namely: cuts in small
$\Delta p_T$ bins can be made to select small $\Delta \tau$ bins in real time that
can exhibit  the various hadronic configurations  of a phase transition process.
Based on the intuition gained in this work, we conjecture that the general features
of our result  are insensitive to the precise validity of this assumption, provided
that $\Delta p_T$ is small enough.  Only by doing the appropriate analysis of the
actual heavy-ion collision data can we assess the plausibility of this assumption
and the feasibility of the whole program discussed here to extract the critical
behavior of quark-hadron phase transition.

It should be noted finally that there are two aspects about the work presented
here.  One is about the theoretical understanding of the properties of the critical
behavior at the quark-hadron PT.  The other is the attempt to find the means to
determine those properties experimentally.  Even if the latter effort fails because
of our incomplete understanding of the actual collision and hadronization process,
the former will remain valid and stand ready for verification by better
phenomenology to come.  Our result on the insensitivity of our proposed measure
to the final-state  effects  provides the encouragement to persist in this difficult
but worthwhile quest.

\begin{center}
\subsubsection*{Acknowledgment}
\end{center}
We are grateful to Drs. Z. Cao and Y. Gao for helpful discussions and assistance.
This work was supported in part by U.S. Department of Energy under
Grant No. DE-FG03-96ER40972.

\newpage
\begin{center}
\section*{Figure Captions}
\end{center}
\begin{description}

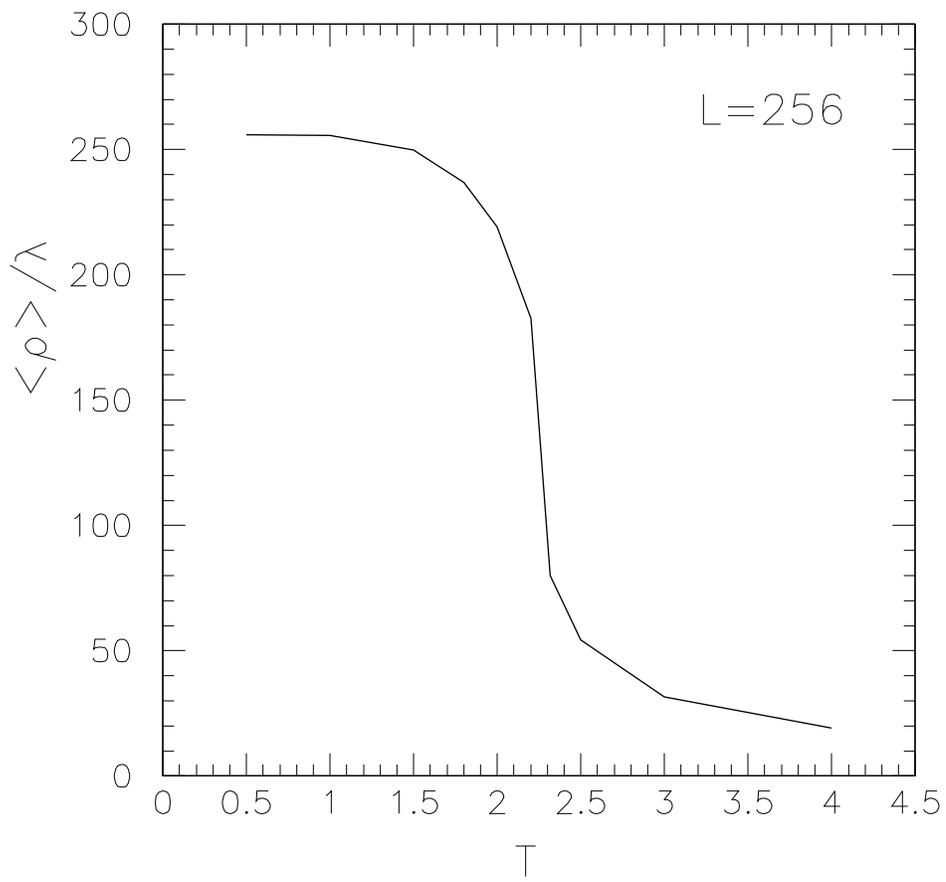
\item[Fig.\ 1]\quad Temperature dependence of the average hadron density in
arbitrary scale.

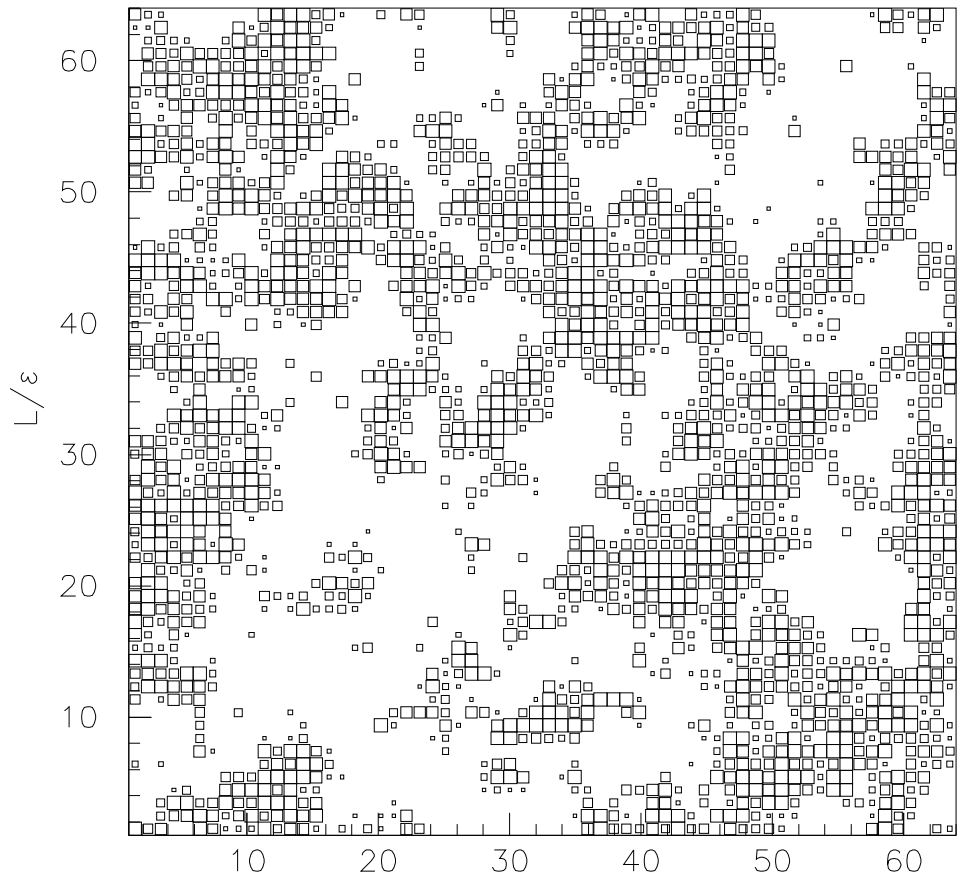
\item[Fig.\ 2]\quad A representative configuration of hadron production in
$\eta$-$\phi$ plane as simulated on the Ising lattice of size $\ell \times \ell$,
where $\ell = L/\epsilon$ is the number of cells  along each dimension. The size of
a square is proportional to the hadron density of a cell.

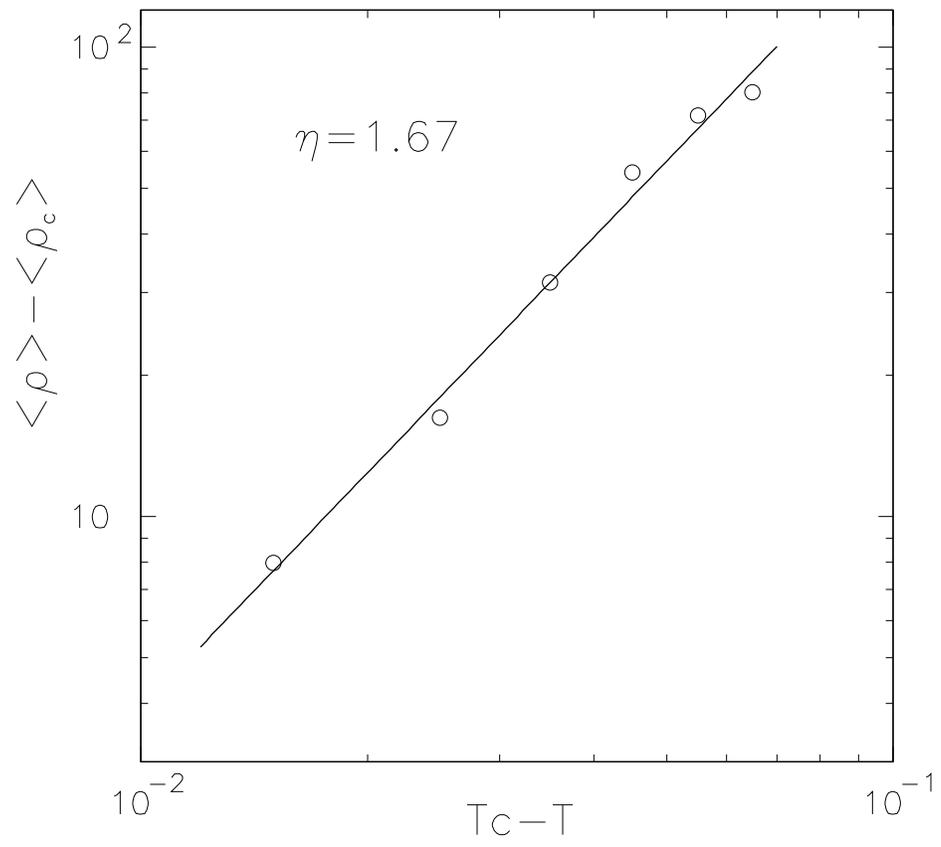
\item[Fig.\ 3]\quad Temperature dependence of average density for $T\leq T_c$.
$\eta$ is the slope of the straightline fit.

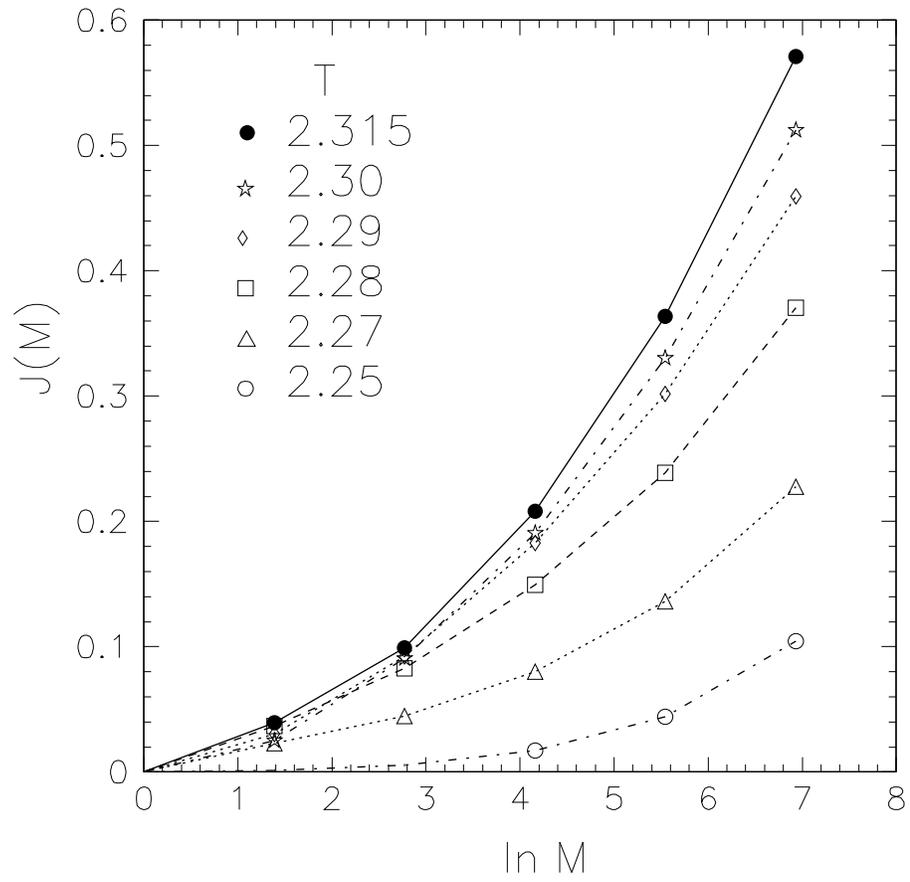
\item[Fig.\ 4]\quad The dependence of $J(M)$ on $M$ for various values of $T$.

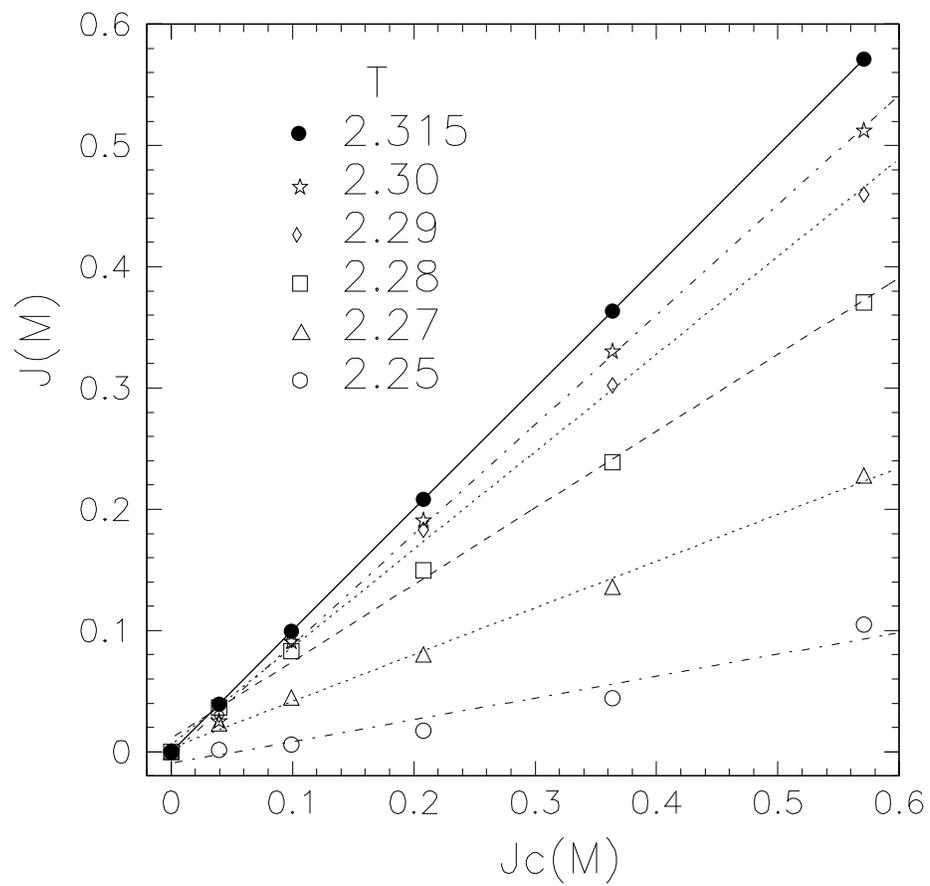
\item[Fig.\ 5]\quad The dependence of $J(M)$ on $J_c(M)$ for various values of
$T$, where $J_c(M)=J(M,T)$ at $T=T_c$.

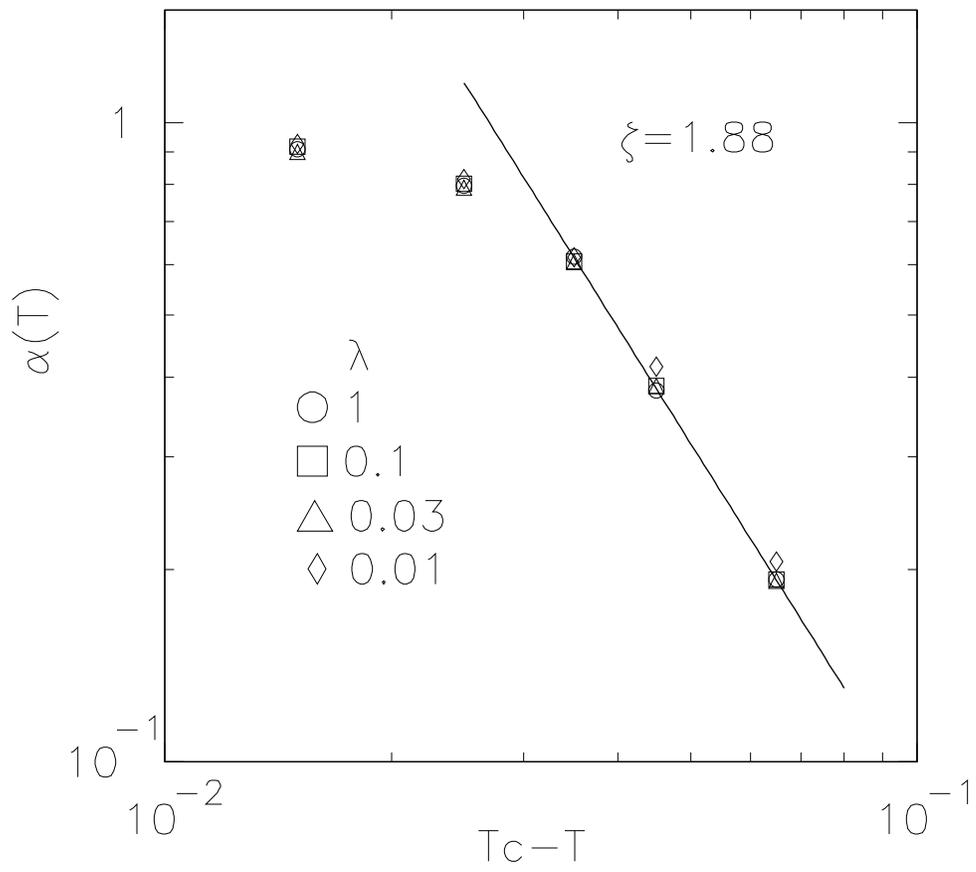
\item[Fig.\ 6]\quad Critical behavior of $\alpha(T)$, where $\alpha(T)$ is defined
in (\ref{14}).  $-\zeta$ is the slope of the straight line.

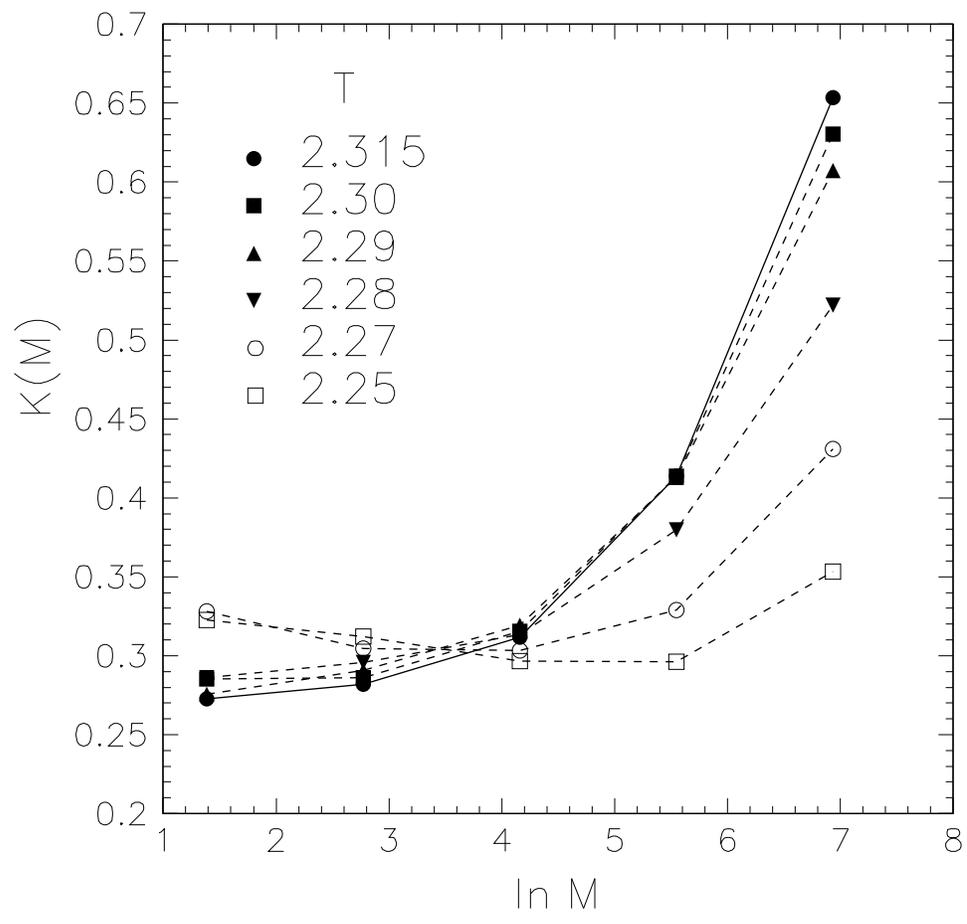
\item[Fig.\ 7]\quad The dependence of $K(M)$ on $M$ for various values of $T$.

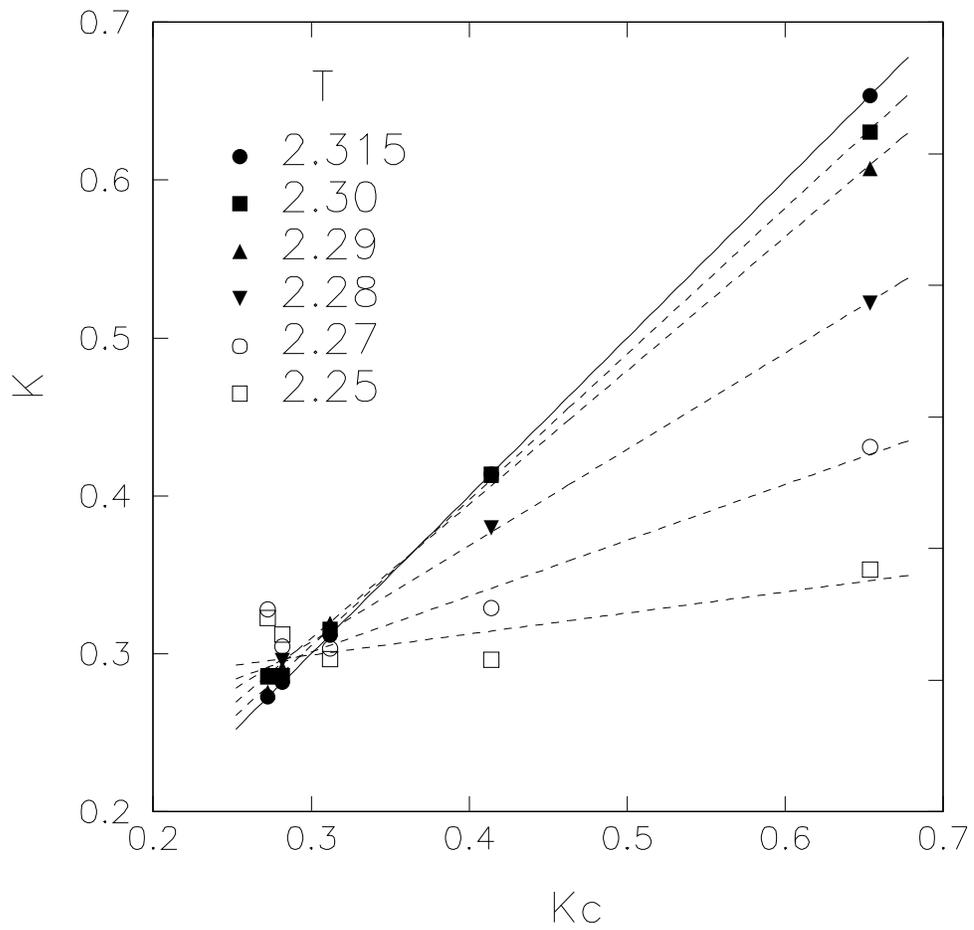
\item[Fig.\ 8]\quad The dependence of $K(M)$ on $K_c(M)$ for various values of
$T$, where $K_c(M)=K(M,T)$ at $T=T_c$.

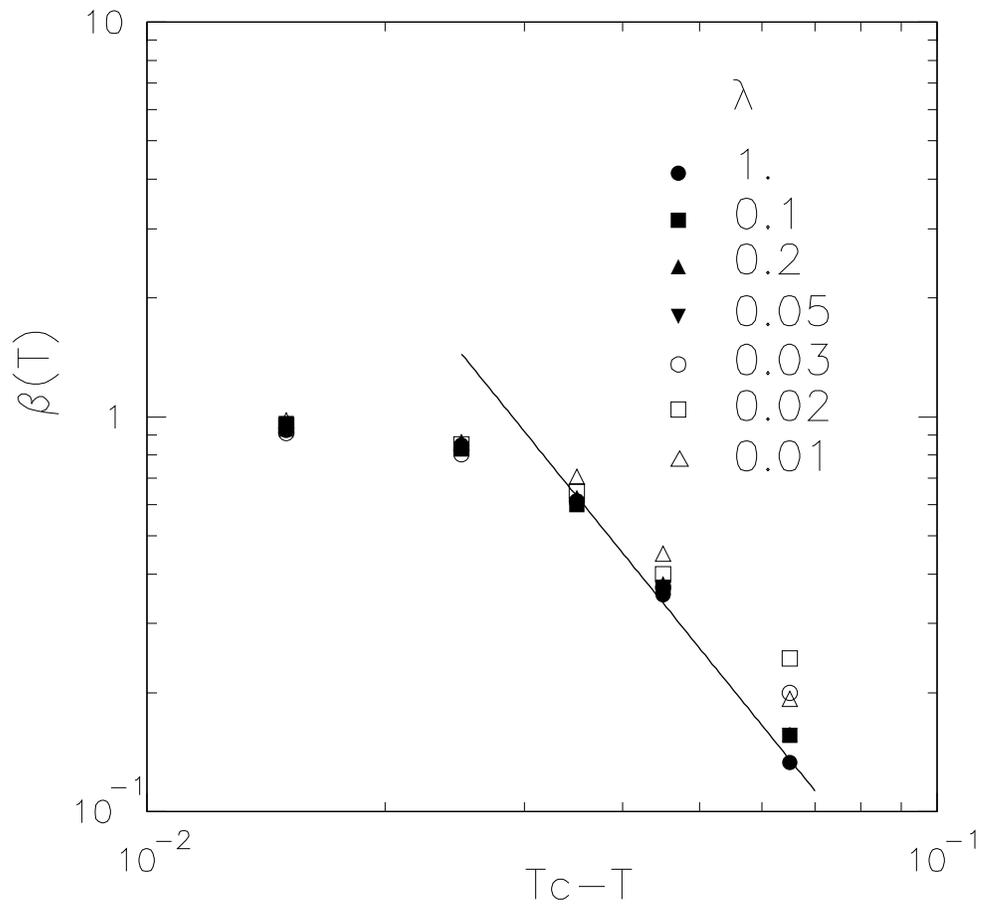
\item[Fig.\ 9]\quad The dependence of $\beta(T)$, defined
in (\ref{27}), on $T$ for various values of $\lambda$.   The slope of the straight
line is
$-\kappa=-2.2$.

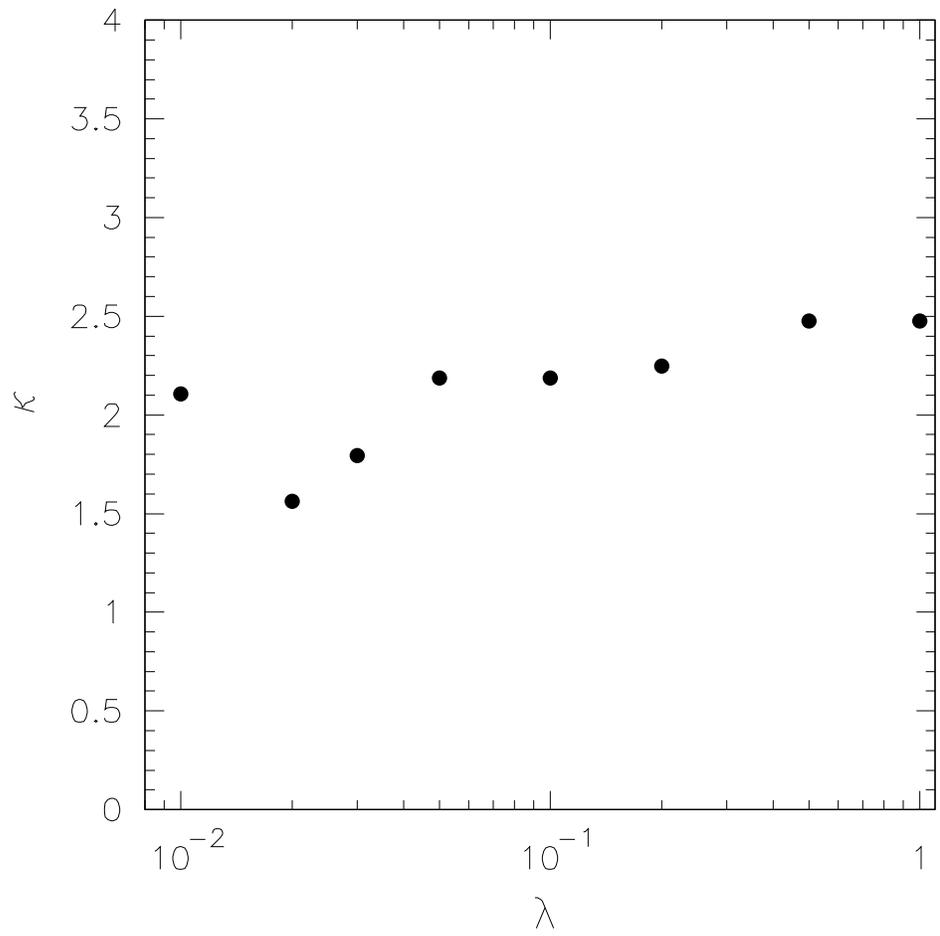
\item[Fig.\ 10]\quad The dependence of the critical exponent $\kappa$ on
$\lambda$.

\item[Fig.\ 11]\quad The direct relationship between $\beta(T)$ and $\alpha(T)$,
when $T$ is varied.

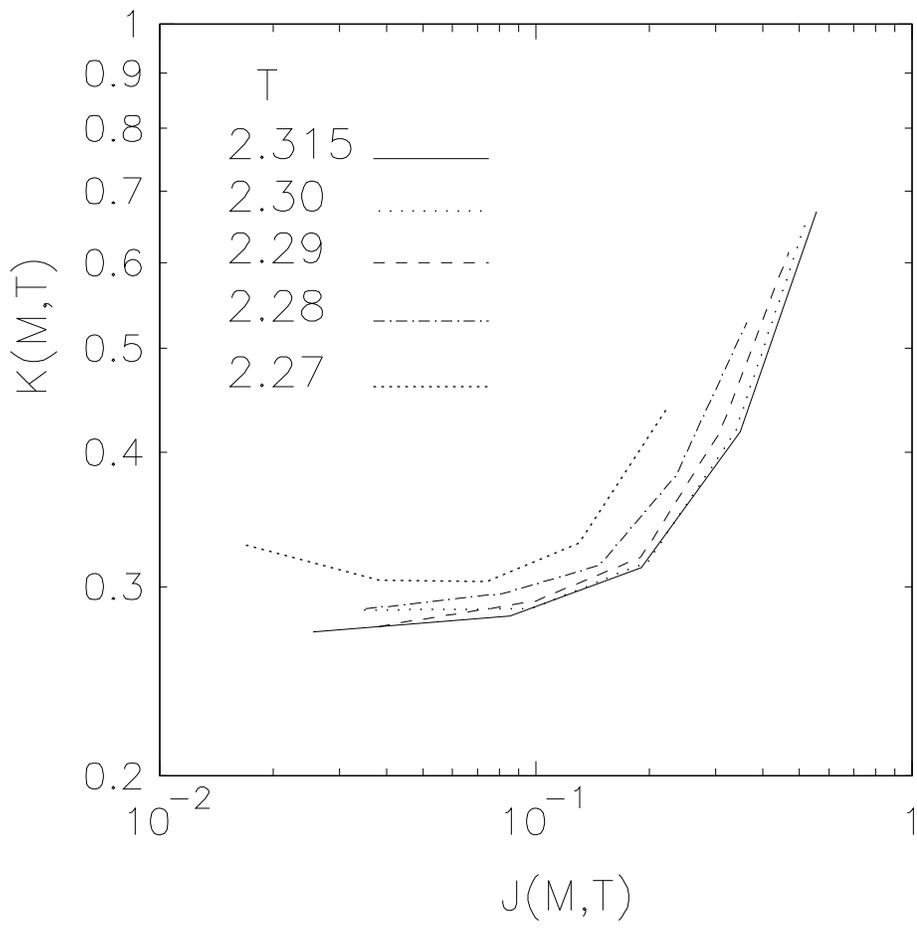
\item[Fig.\ 12]\quad The direct relationship between $K(M,T)$ and $J(M,T)$,
when $M$ is varied but $T$ fixed.

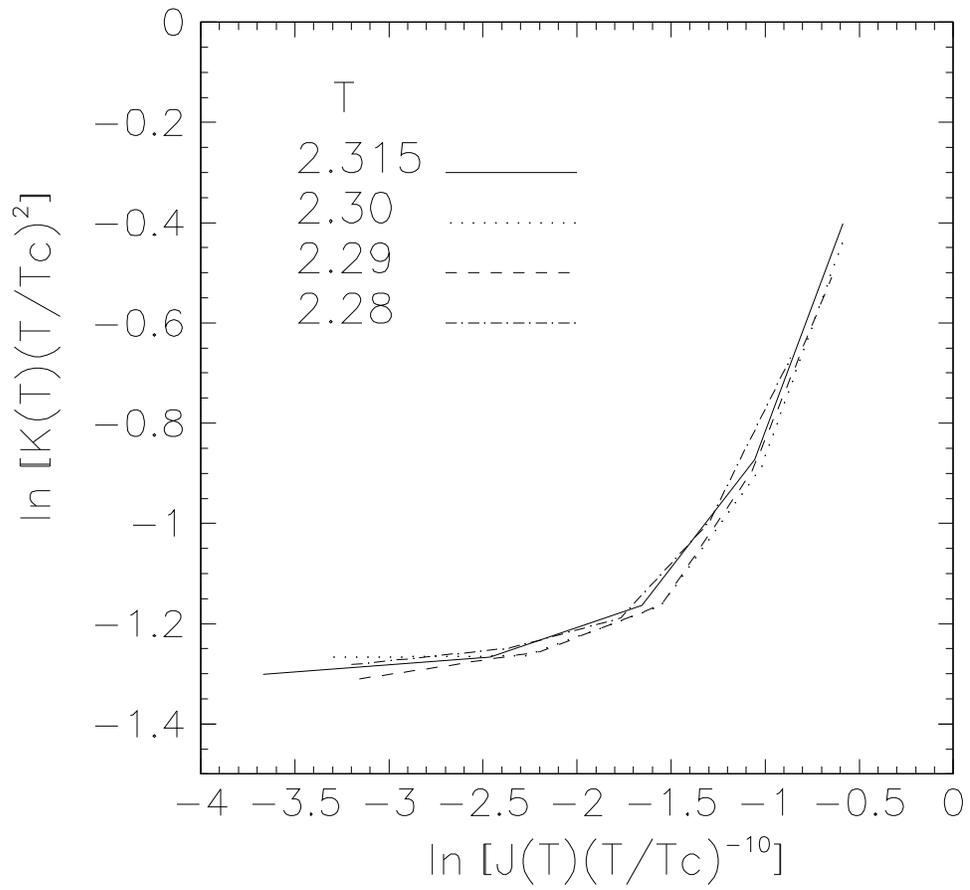
\item[Fig.\ 13]\quad Same as Fig.\ 12 but in rescaled variables.

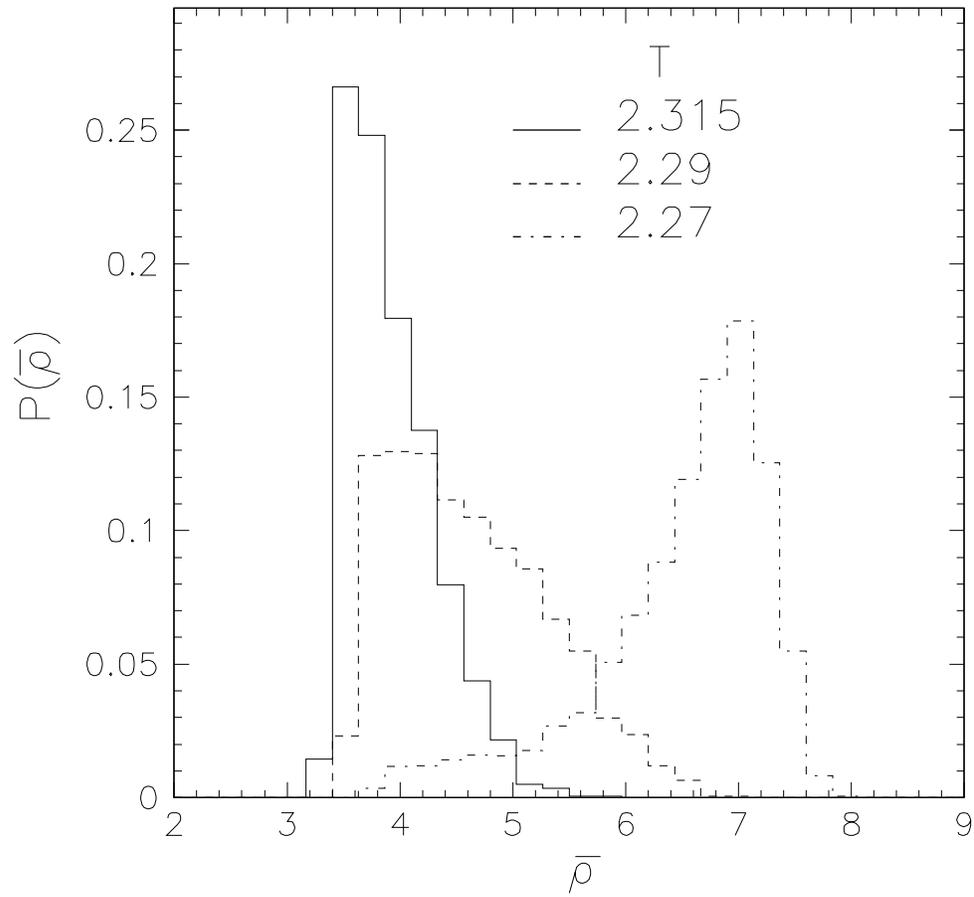
\item[Fig.\ 14]\quad The probability distribution of the average
density $\bar\rho$ in a configuration for three fixed $T$, generated by $10^4$
configurations at each $T$.   The scale of $\bar\rho$ is arbitrary.

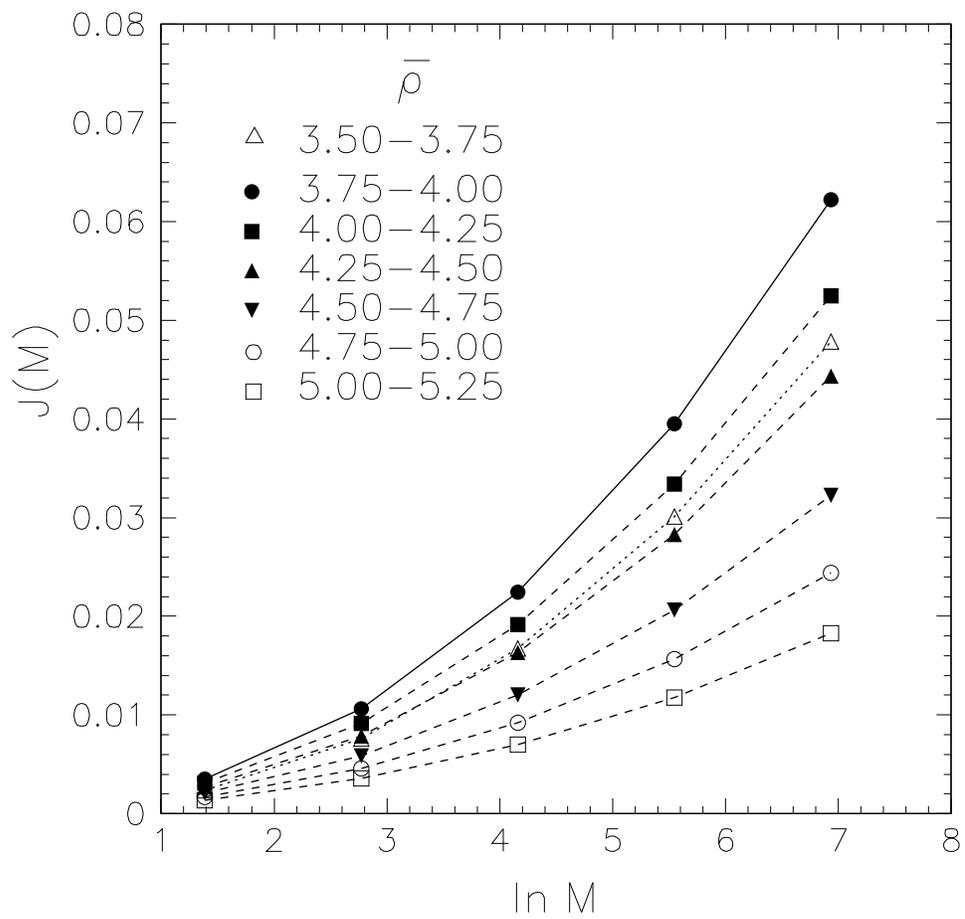
\item[Fig.\ 15]\quad The dependence of $J(M, \bar\rho)$ on $M$ for various
values of $\bar\rho$.

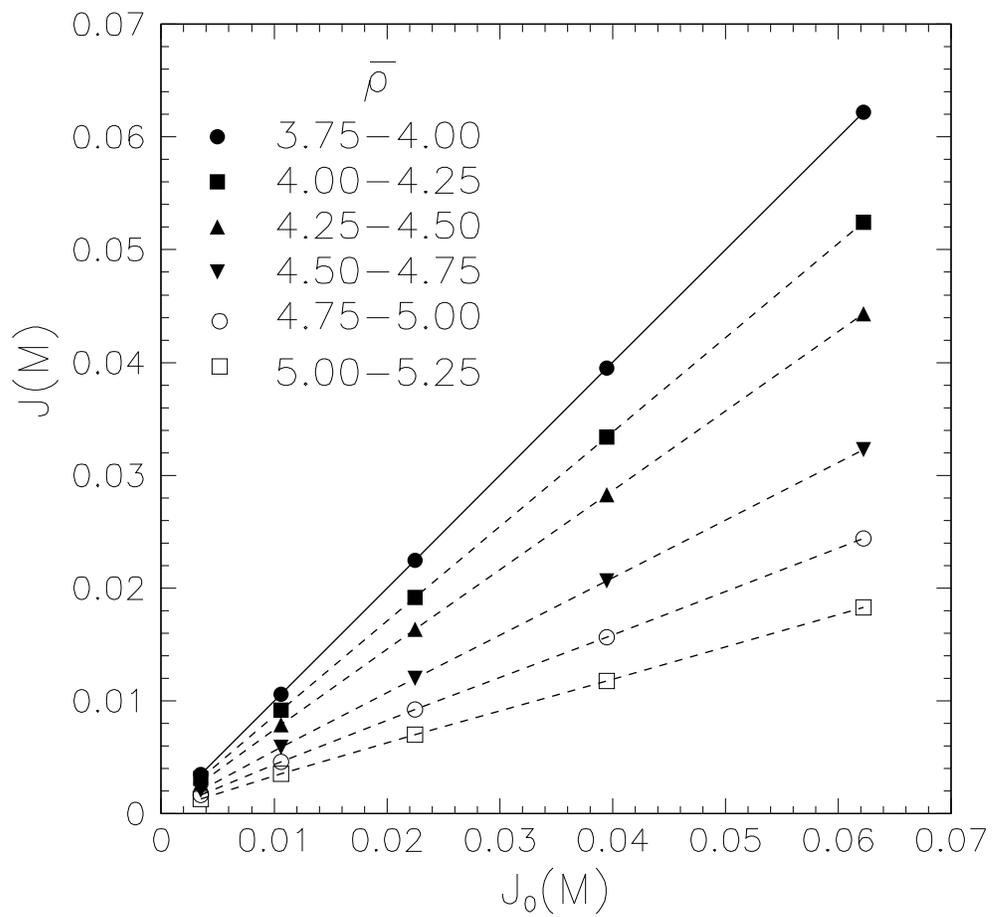
\item[Fig.\ 16]\quad The dependence of $J(M, \bar\rho)$ on $J_0(M)$ for various
values of $\bar\rho$, where $J_0(M)=J(M,\bar\rho)$ at $\bar\rho=\bar\rho_0$.

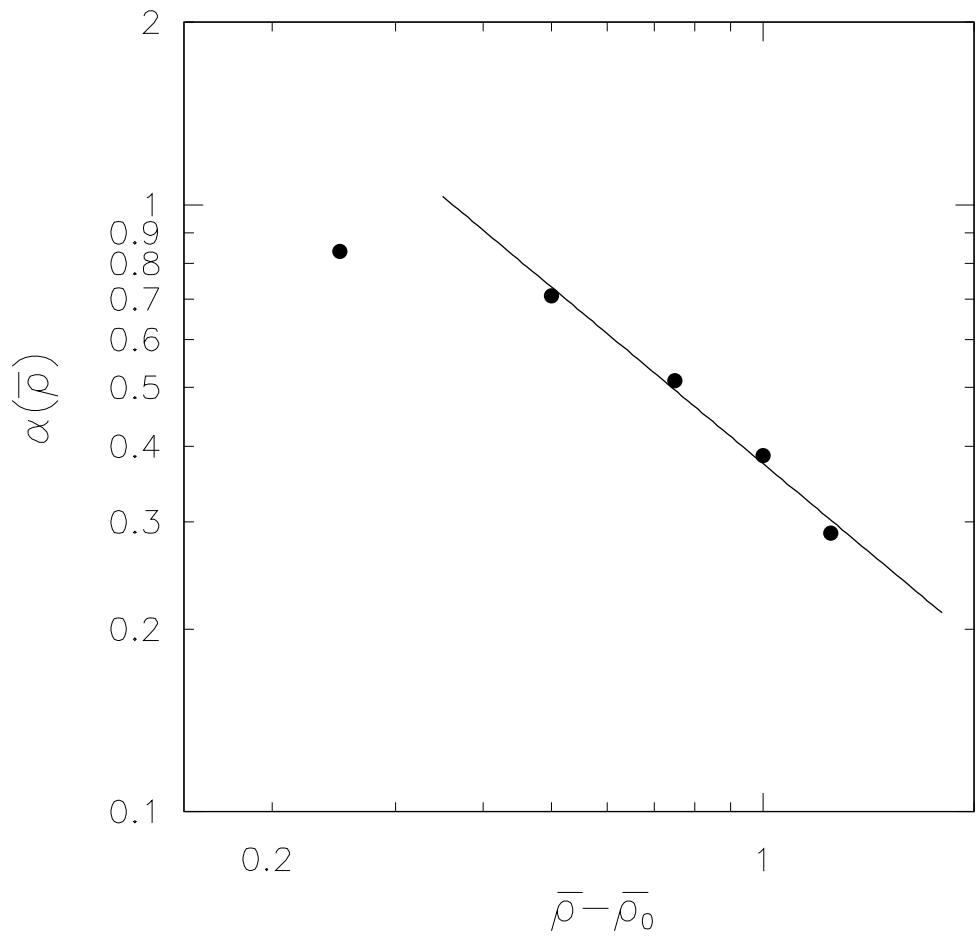
\item[Fig.\ 17]\quad Critical behavior of $\alpha(\bar\rho)$, where
$\alpha(\bar\rho)$ is the slope of a straightline fit in Fig.\ 16.

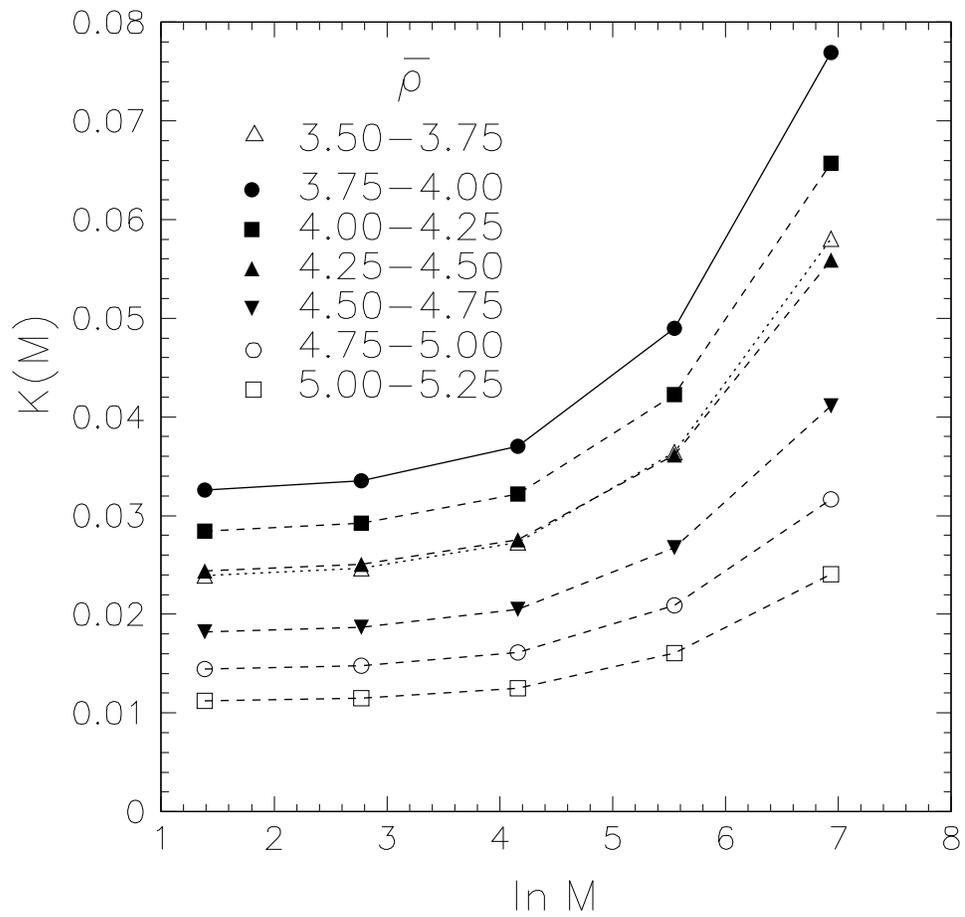
\item[Fig.\ 18]\quad The dependence of $K(M, \bar\rho)$ on $M$ for various
values of $\bar\rho$.

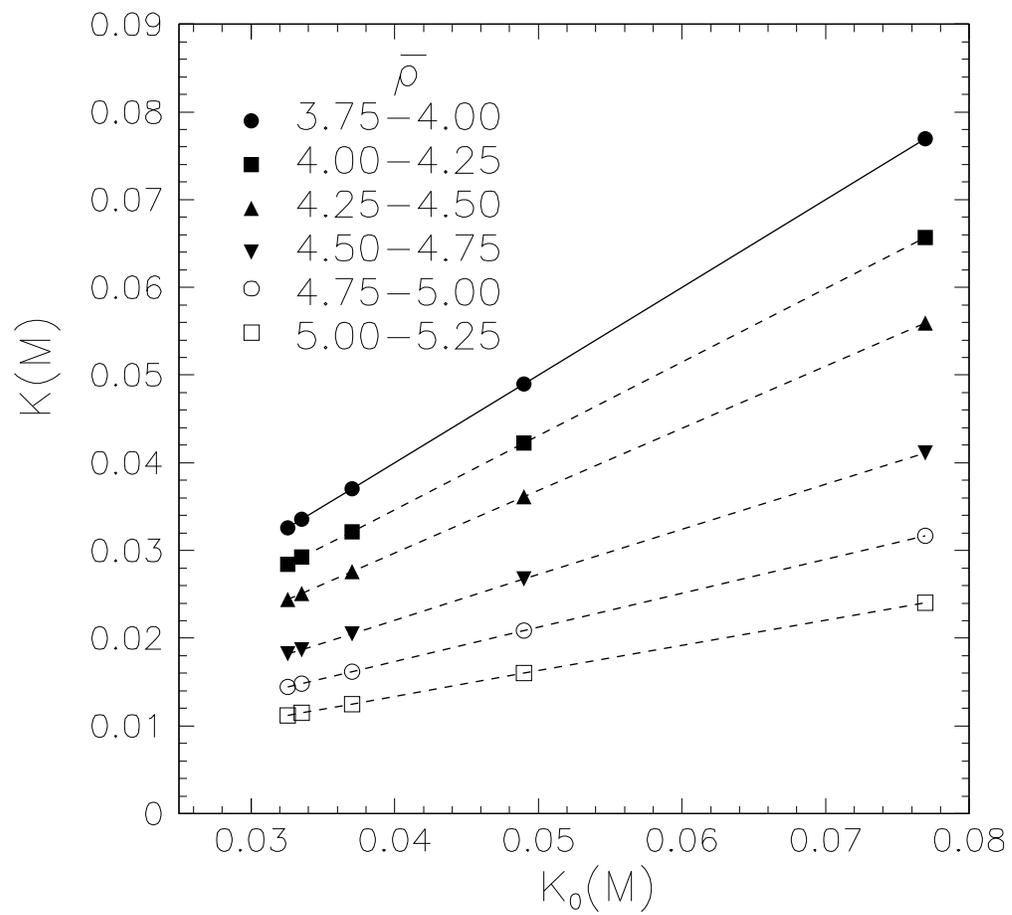
\item[Fig.\ 19]\quad The dependence of $K(M, \bar\rho)$ on $K_0(M)$ for various
values of $\bar\rho$, where $K_0(M)=K(M,\bar\rho)$ at $\bar\rho=\bar\rho_0$.

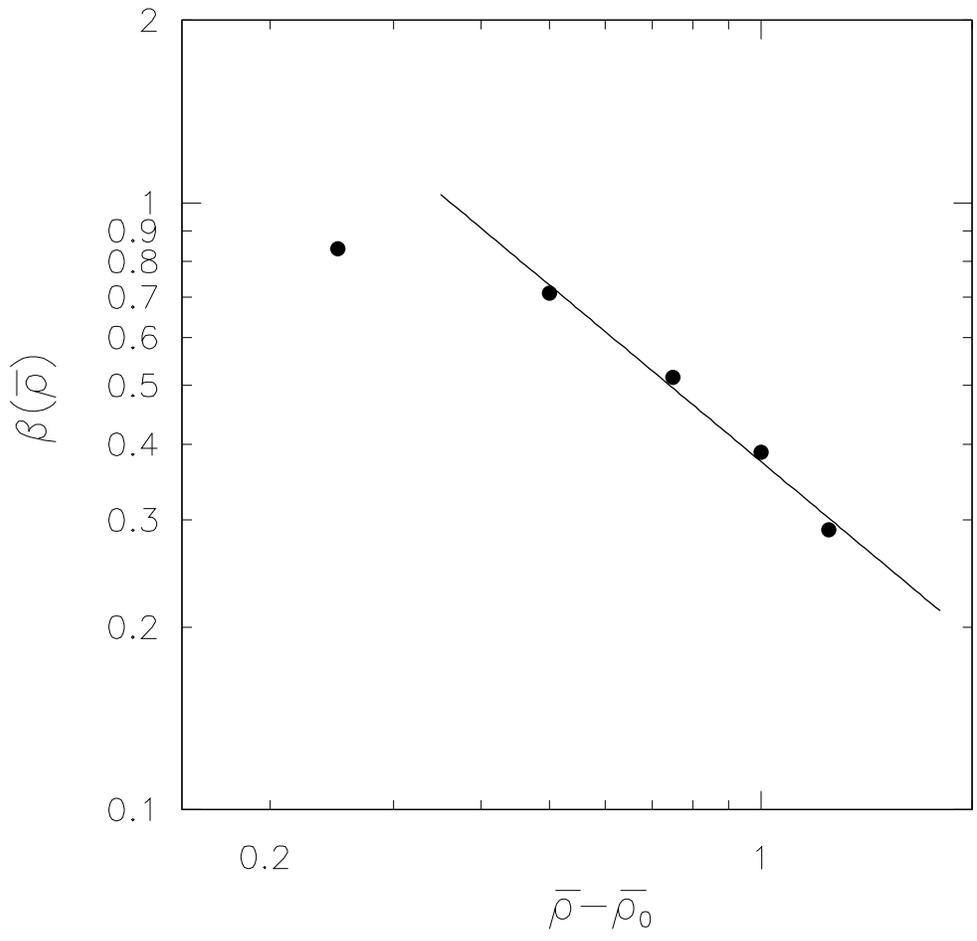
\item[Fig.\ 20]\quad Critical behavior of $\beta(\bar\rho)$, where
$\beta(\bar\rho)$ is the slope of a straightline fit in Fig.\ 19.

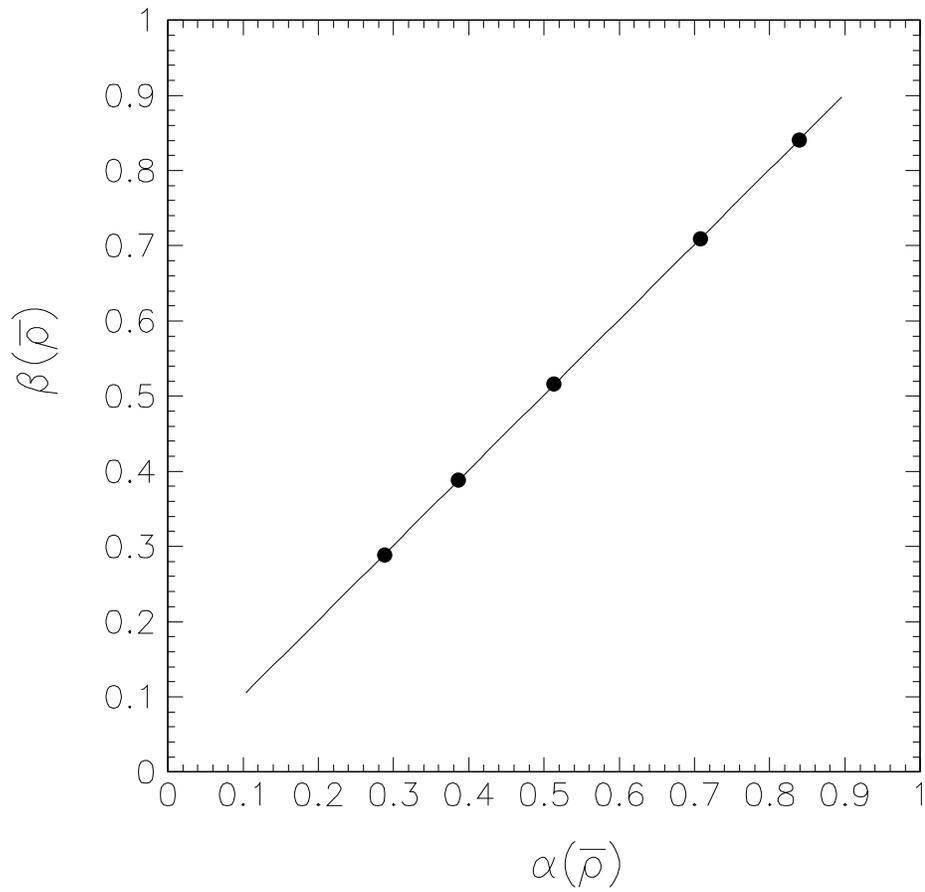
\item[Fig.\ 21]\quad The direct relationship between $\beta(\bar\rho)$ and
$\alpha(\bar\rho)$, when $\bar\rho$ is varied.

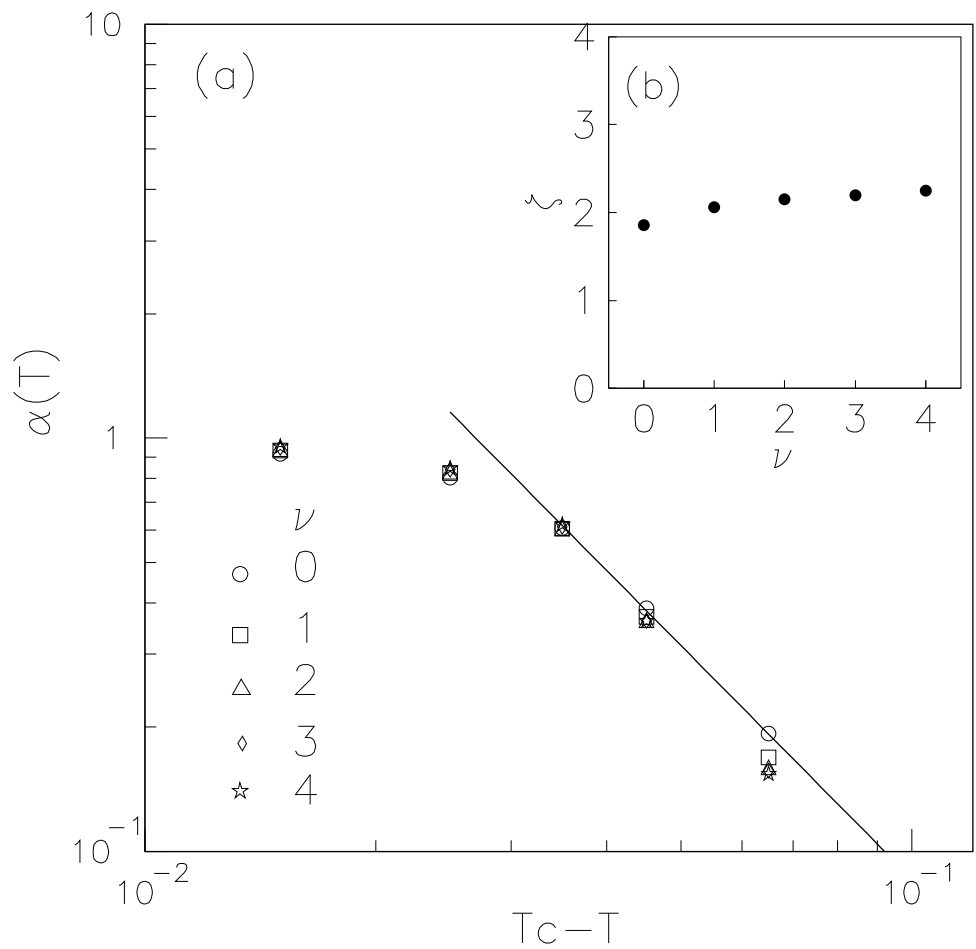
\item[Fig.\ 22]\quad The effect of randomization on (a) $\alpha(T)$ and (b)
$\zeta$, where $\nu$ is the number of steps of random walk.

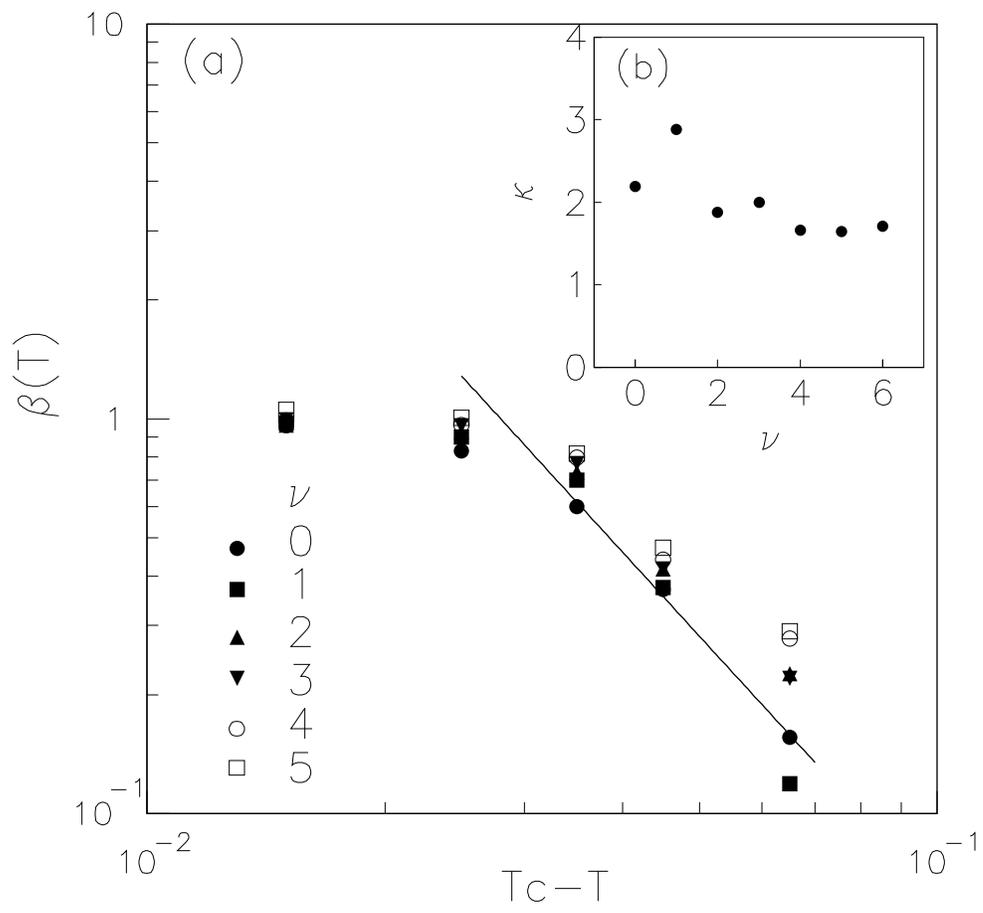
\item[Fig.\ 23]\quad The effect of randomization on (a) $\beta(T)$ and (b)
$\kappa$, where $\nu$ is the number of steps of random walk.

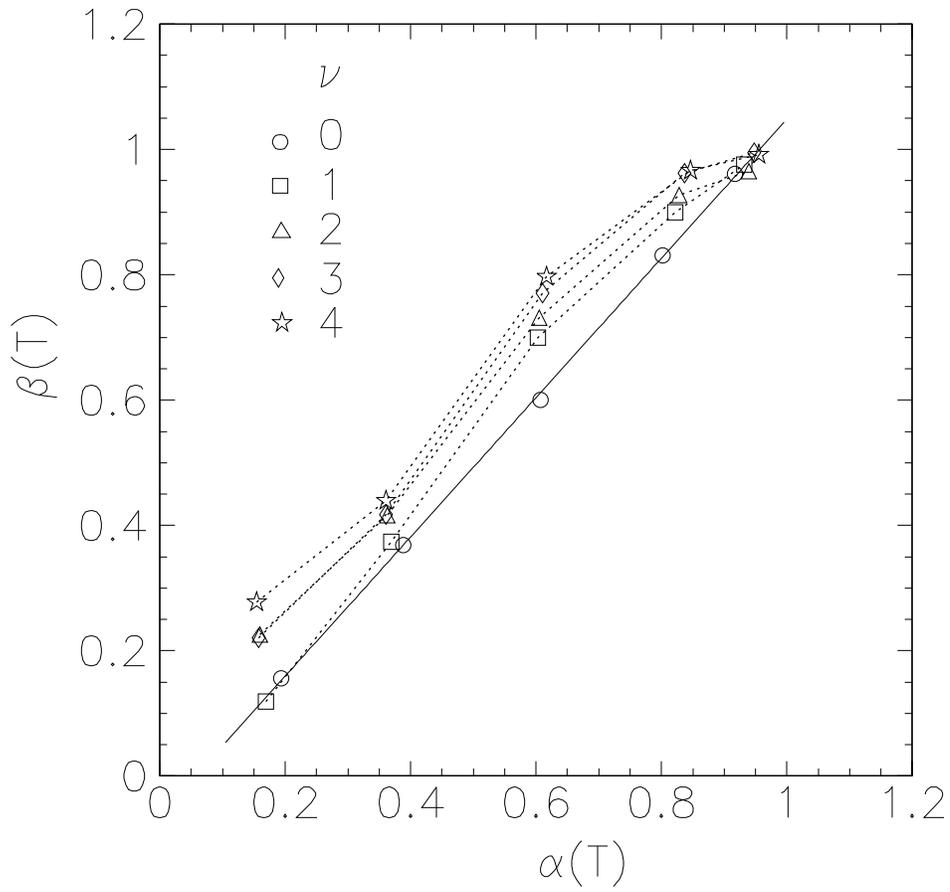
\item[Fig.\ 24]\quad The effect of randomization on the relationship between
$\beta(T)$ and $\alpha(T)$.

\end{description}

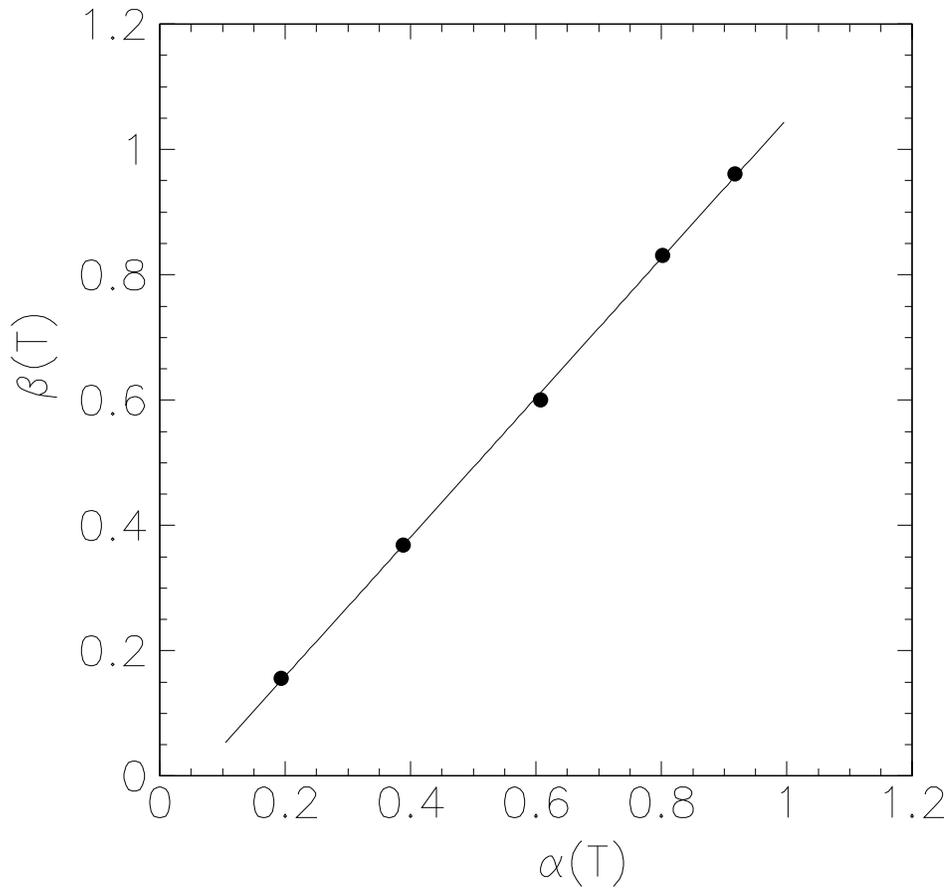
\end{document}